\documentclass[10pt,english,prb,amsmath,amssymb,prb,showkeys,superscriptaddress,twocolumn,showpacs,floatfix]{revtex4-1}
\usepackage{graphicx}
\usepackage{dcolumn}
\usepackage[colorlinks=true,linkcolor=blue ,citecolor=blue,urlcolor=blue]{hyperref}
\usepackage{multirow}
\usepackage[usenames,dvipsnames]{xcolor}
\usepackage{soul}
\usepackage{braket}
\usepackage{bm}
\usepackage{nicefrac}
\usepackage{siunitx}
\usepackage{color,transparent}
\usepackage[utf8]{inputenc}
\usepackage{upgreek}

\newcommand{\HH}{\mathcal{H}}

\newcommand{\iu}{\mathrm{i}}

\newcommand{\muB}{\mu_{\text{B}}}

\newcommand{\VEC}[1]{\mathbf{#1}}

\newcommand{\MC}[1]{\mathcal{#1}}

\DeclareSIUnit\ML{ML}
\DeclareSIUnit\MLs{MLs}
\DeclareSIUnit\meVA{meV\angstrom^2}

\newcommand{\icol}[1]{ \left(\begin{smallmatrix}#1\end{smallmatrix}\right) }

\begin{document}

\title{Spin--waves in the collinear antiferromagnetic phase of \texorpdfstring{\uppercase{M}\lowercase{n}\textsubscript{5}\uppercase{S}\lowercase{i}\textsubscript{3}}{Mn5Si3}}

\author{F. J. dos Santos}
\email{f.dos.santos@fz-juelich.de (currently based at EPFL: flaviano.dossantos@epfl.ch)}
\affiliation{Peter Gr\"unberg Institut and Institute for Advanced Simulations, Forschungszentrum J\"ulich $\&$ JARA, D-52425 J\"ulich, Germany}
\affiliation{Department of Physics, RWTH Aachen University, 52056 Aachen, Germany}
\author{N. Biniskos}
\email{n.biniskos@fz-juelich.de}
\affiliation{Forschungszentrum J\"ulich GmbH, J\"ulich Centre for Neutron Science at MLZ, Lichtenbergstr. 1, 85748 Garching, Germany}
\author{S. Raymond}
\affiliation{Universit\'e Grenoble Alpes, CEA, IRIG, MEM, 38000 Grenoble, France}
\author{K. Schmalzl}
\affiliation{Forschungszentrum J\"ulich GmbH, J\"ulich Centre for Neutron Science at ILL, 71 avenue des Martyrs, 38000 Grenoble, France}
\author{M. dos Santos Dias}
\affiliation{Peter Gr\"unberg Institut and Institute for Advanced Simulations, Forschungszentrum J\"ulich $\&$ JARA, D-52425 J\"ulich, Germany}
\author{P. Steffens}
\affiliation{Institut Laue-Langevin, 71 avenue des Martyrs, 38000 Grenoble, France}
\author{J. Persson}
\affiliation{Forschungszentrum J\"ulich GmbH, J\"ulich Centre for Neutron Science (JCNS-2) and Peter Gr\"unberg Institut (PGI-4), JARA-FIT,  52425 J\"ulich, Germany}
\author{S. Bl\"ugel}
\affiliation{Peter Gr\"unberg Institut and Institute for Advanced Simulations, Forschungszentrum J\"ulich $\&$ JARA, D-52425 J\"ulich, Germany}
\author{S. Lounis}
\affiliation{Peter Gr\"unberg Institut and Institute for Advanced Simulations, Forschungszentrum J\"ulich $\&$ JARA, D-52425 J\"ulich, Germany}
\affiliation{Faculty of Physics, University of Duisburg-Essen, 47053 Duisburg, Germany}
\author{T. Br\"uckel}
\affiliation{Forschungszentrum J\"ulich GmbH, J\"ulich Centre for Neutron Science (JCNS-2) and Peter Gr\"unberg Institut (PGI-4), JARA-FIT,  52425 J\"ulich, Germany}
\date{\today}

\begin{abstract}
By combining two independent approaches, inelastic neutron scattering measurements and density functional theory calculations, we study the spin--waves in the collinear antiferromagnetic phase (AFM2) of $\textnormal{Mn}_5\textnormal{Si}_3$.
We obtain its magnetic ground--state properties and electronic structure.
This study allowed us to determine the dominant magnetic exchange interactions and magnetocrystalline anisotropy in the AFM2 phase of $\textnormal{Mn}_5\textnormal{Si}_3$.
Moreover, the evolution of the spin excitation spectrum is investigated under the influence of an external magnetic field perpendicular to the anisotropy easy--axis. The low energy magnon modes show a different magnetic field dependence which is a direct consequence of their different precessional nature. 
Finally, possible effects related to the Dzyaloshinskii--Moriya interaction are also considered.

\end{abstract}

\date{\today}

\maketitle

\section{Introduction}

    The study of magnetism at a microscopic level can lead to designing cutting--edge technological applications for data process and storage, information transmission, and magnetic refrigeration. In recent years, antiferromagnetic (AFM) materials have attracted great interest in the research field of spintronics~\cite{Baltz_2018}. Bulk $\textnormal{Mn}_5\textnormal{Si}_3$ is an AFM intermetallic compound that is hosting rich physics. Its interesting properties, such as the complex magnetic structure~\cite{brown_low-temperature_1992}, the anomalous Hall effect~\cite{surgers_large_2014}, and the inverse magnetocaloric effect~\cite{biniskos_spin_2018} have been attributed to an instability of the Mn magnetic moments. It is also worth mentioning that in nanoparticle~\cite{das_mn5si3_2016} and nanowire~\cite{sun_millimeters_2020} form, $\textnormal{Mn}_5\textnormal{Si}_3$ is considered to have great potential in future electronic and spintronic devices. Despite the intense research activity over the past decades~\cite{brown_low-temperature_1992,surgers_large_2014,biniskos_spin_2018,gottschilch_study_2012,brown_antiferromagnetism_1995,surgers_switching_2017,vinokurova_magnetic_1990,luccas_magnetic_2019,das_observation_2019,silva_magnetic_2002,Alkanani_1995,Irizawa_2002,Suergers_2016}, many open questions remain regarding the minimal magnetic model Hamiltonian, the role of the spin fluctuations in the magnetically ordered phases and which Mn site is responsible for them. To address some of these questions, in the present study, we perform inelastic neutron scattering experiments and apply first--principles calculations. 

   The crystal and magnetic structure of bulk Mn$_{5}$Si$_{3}$ has been established by neutron diffraction measurements~\cite{gottschilch_study_2012,brown_antiferromagnetism_1995,brown_low-temperature_1992} and the magnetic phase diagram as a function of temperature and magnetic field has been extensively studied by magnetization and electrical transport measurements~\cite{surgers_switching_2017,vinokurova_magnetic_1990,luccas_magnetic_2019,das_observation_2019,songlin_magnetic_2002}. In the paramagnetic (PM) state, $\textnormal{Mn}_5\textnormal{Si}_3$ crystallizes in the hexagonal space group $P6_{3}/mcm$ with two distinct crystallographic positions for Mn atoms (sites Mn1 and Mn2)~\cite{gottschilch_study_2012} and undergoes two successive first order phase transitions towards antiferromagnetic phases, which occur at $T_{N_{2}}\approx$ 100\,K (AFM2) and $T_{N_{1}}\approx$ 66\,K (AFM1), respectively~\cite{songlin_magnetic_2002}. The electric resistivity shows a metallic behavior with two anomalies corresponding to these phase transitions~\cite{vinokurova_magnetic_1990}.

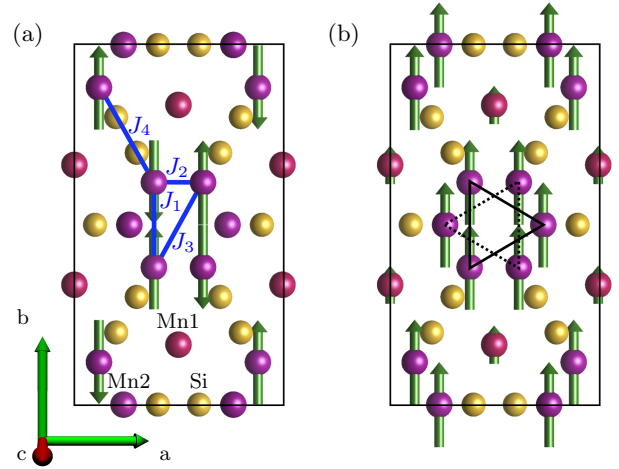
\begin{figure}[tb]
\setlength{\unitlength}{1cm} 
\newcommand{\boxsize}{0.3}

  \begin{picture}(7,6)
    
    \put(0.0, 0.0){    
        \put(0.0, 0.0){ \includegraphics[width=3.3 cm,trim={0 0 0 0},clip=true]{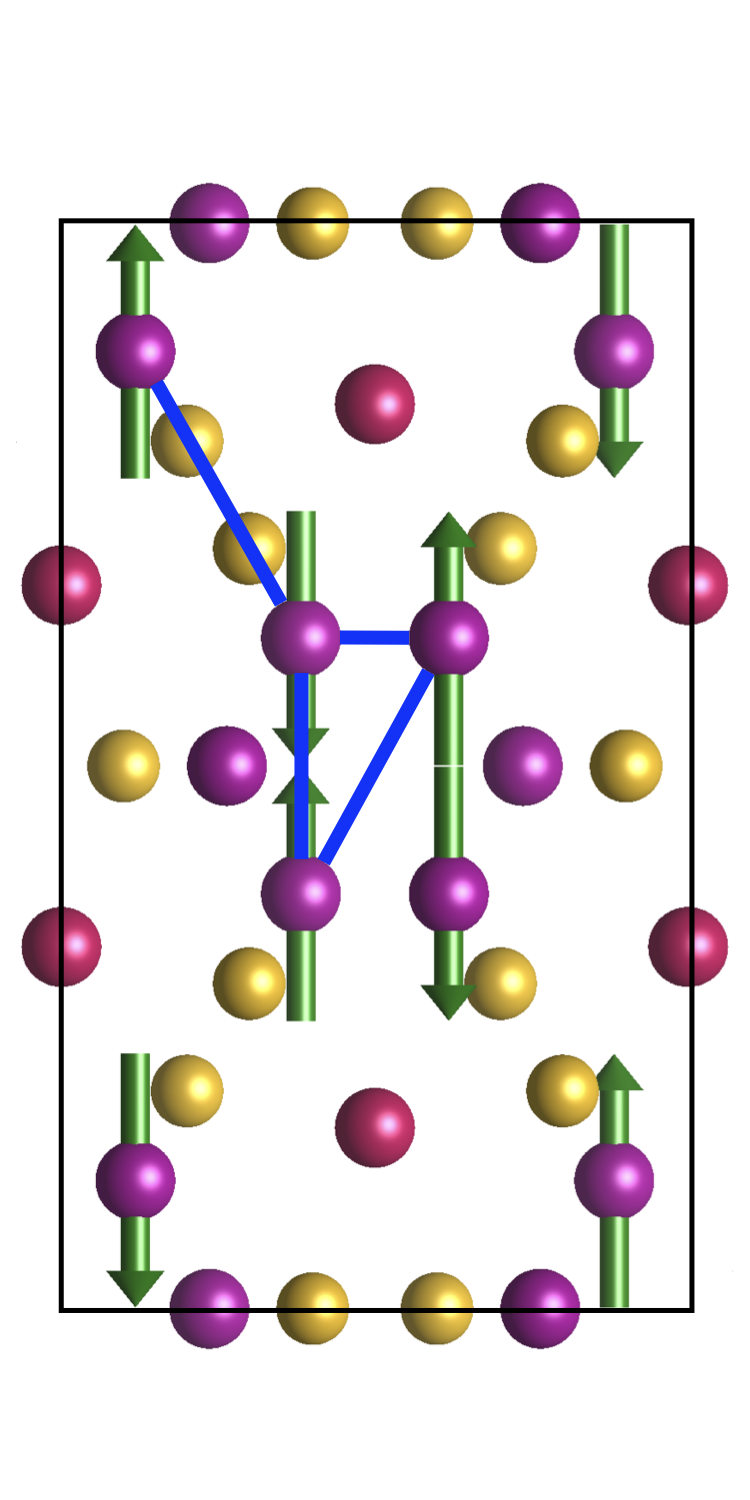} }
        \put(4.2, 0.0){ \includegraphics[width=3.3 cm,trim={0 0 0 0},clip=true]{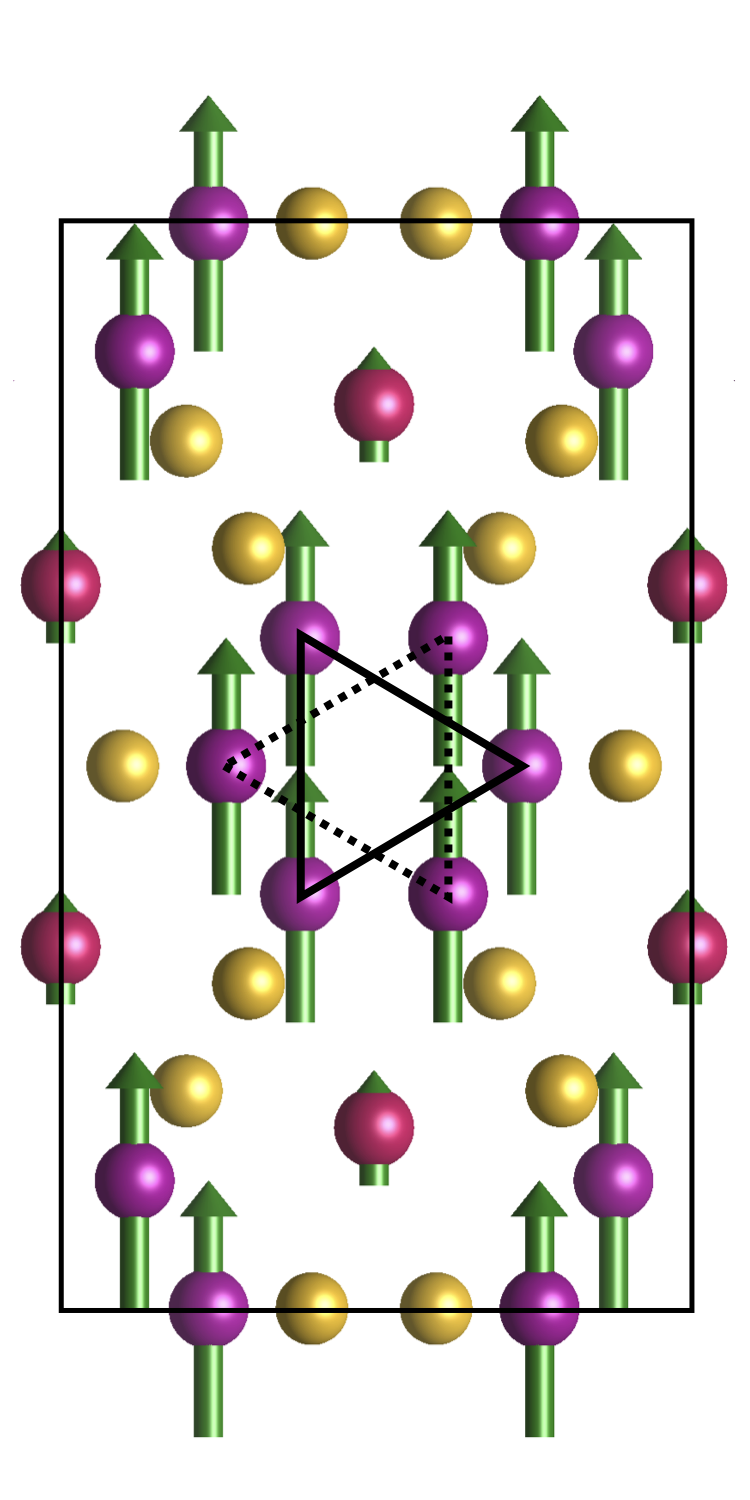} }
    
        \put(-0.5, 5.6){ \makebox(\boxsize,\boxsize){(a)} } 
        \put( 3.7, 5.6){ \makebox(\boxsize,\boxsize){(b)} } 

        \put( 1.80, 1.05){ \scalebox{0.9}{\makebox{Si}} }
        \put( 0.70, 1.05){ \scalebox{0.9}{\makebox{Mn2}} }
        \put( 1.36, 1.85){ \scalebox{0.9}{\makebox{Mn1}} } 
        {\color{blue}
        \put( 0.97, 4.45){ \scalebox{1.0}{\makebox{$J_4$}} }
        \put( 1.47, 3.90){ \scalebox{1.0}{\makebox{$J_2$}} }
        \put( 1.39, 3.45){ \scalebox{1.0}{\makebox{$J_1$}} }
        \put( 1.55, 2.90){ \scalebox{1.0}{\makebox{$J_3$}} }
        }
        \put(-0.5,0.0) {
            \put( 0.05, 0 ){ \includegraphics[width=1.7cm,trim={30cm 29cm 10cm  10cm  },clip=true]{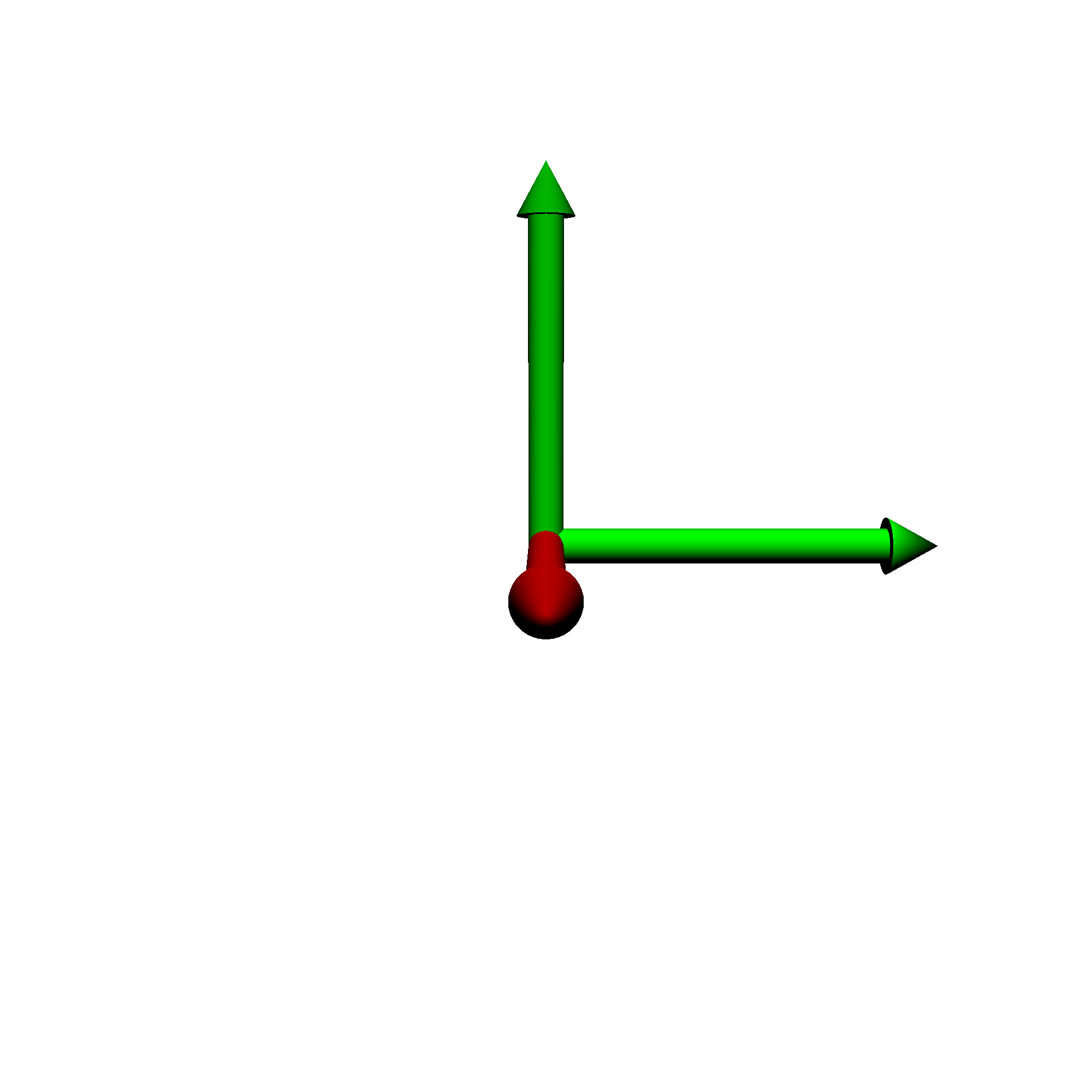} }
            \put( 1.9, 0.1){ \scalebox{0.9}{\makebox{a}} }
            \put(   0, 1.9){ \scalebox{0.9}{\makebox{b}} }
            \put(   0, 0.1){ \scalebox{0.9}{\makebox{c}} }
        }
    }
  \end{picture}

  \caption{\label{fig:Capture1}
  (a) Projection of the structure of $\textnormal{Mn}_5\textnormal{Si}_3$ in the AFM2 phase in the $ab$--plane of the orthorhombic cell according to single--crystal neutron diffraction data~\cite{brown_antiferromagnetism_1995}. The blue solid lines indicate the relevant exchange interactions used in the Heisenberg Hamiltonian.
  (b) Predicted metastable FM phase from first--principles calculations (see details in text). The two triangles indicate Mn2 atoms located in different planes.
  }
\end{figure}

   At $T_{N_{2}}\approx$ 100\,K (AFM2), the crystal structure changes from hexagonal to orthorhombic with space group $Ccmm$ and Mn2 divides into two sets of inequivalent positions~\cite{brown_antiferromagnetism_1995}. In the orthorhombic cell magnetic reflections follow the condition $h+k$ odd and the magnetic propagation vector is $\bm\kappa = (0, 1, 0)$. In the AFM2 phase, the Mn1 and one--third of the Mn2 atoms have no ordered moments and the remaining Mn2 atoms have their magnetic moments of magnitude 1.48(1)\,$\mu_{B}$ aligned almost parallel and antiparallel to the $b$--axis of the orthorhombic cell~\cite{brown_antiferromagnetism_1995} (see Fig.~\ref{fig:Capture1}(a)). At a lower temperature, at $T_{N_{1}}\approx$ 66\,K (AFM1), a structural distortion occurs to an orthorhombic cell without inversion symmetry (space group $Cc2m$)~\cite{brown_low-temperature_1992}. The magnetic moments reorient in a highly noncollinear and noncoplanar arrangement, while the propagation vector remains the same. Mn1 atoms acquire a magnetic moment of magnitude 1.20(5)\,$\mu_{B}$ and still one--third of the Mn2 atoms have no ordered moments, just as in the AFM2 phase. The rest of the Mn2 atoms carry a moment of 2.30(9) and 1.85(9)\,$\mu_{B}$, depending on their site.

\section{Experimental part}

\subsection{Experimental details}

   The Mn$_{5}$Si$_{3}$ single crystal was grown by the Czochralski method~\cite{biniskos_spin_2018}. The sample with a mass of about 7\,g was mounted on an aluminum sample holder and was oriented in the [100]/[010] scattering plane of the orthorhombic symmetry. Inelastic neutron scattering (INS) measurements were carried out on the cold triple--axis spectrometers (TAS) IN12~\cite{schmalzl_upgrade_2016} and ThALES at the Institut Laue Langevin (ILL) in Grenoble, France. Both TAS were setup in W configuration and inelastic scans were performed with constant $\VEC{k}_f$, where $\VEC{k}_f$ is the wave--vector of the scattered neutron beam.
   
   Unpolarized INS measurements were performed at IN12 and focusing setups were employed. The spectrometer was equipped with pyrolytic graphite (PG(002)) crystals as the monochromator and analyzer and 40'--open--open collimations were installed. Higher--order contamination was removed using a velocity selector (VS) before the monochromator and a beryllium (Be) filter in the scattered neutron beam. The sample was cooled below room temperature with a $^4$He flow cryostat. Spin dynamics investigations with unpolarized neutrons under the magnetic field were carried out using a 10\,T vertical field magnet. For these measurements, the Be filter was removed. The single crystal was cooled down from the PM state to $T = 80$\,K (AFM2 phase) without the presence of an external magnetic field. The field was applied along the $c$--axis of the orthorhombic symmetry of Mn$_{5}$Si$_{3}$ and the spectra were collected with increasing field strength. 
   
   At ThALES longitudinal polarization analysis (LPA) was performed using the CRYOPAD device~\cite{regnault_spherical_2004} to guide and orient the neutron beam polarization with a strictly zero magnetic field in the sample position. The TAS was equipped with polarizing Heusler (Cu$_{2}$MnAl(111)) crystals as the monochromator and analyzer. A flipping ratio of 14 was determined from measurements in a graphite sample. Fully focusing setups were employed and higher--order contamination was removed using a VS and a Be filter before the monochromator and in the scattered neutron beam, respectively. Inelastic scans were performed with a constant $\VEC{k}_f$ of 1.1\,{\AA}$^{-1}$. For the polarized INS experiments the common Cartesian coordinate system was used~\cite{chatterji_neutron_2006}: the $x$--axis parallel to the scattering vector $\VEC Q$~\cite{comment}, the $y$--axis perpendicular to $\VEC Q$ in the scattering plane and the $z$--axis perpendicular to the scattering plane.
   
\subsection{Unpolarized INS measurements}

\begin{figure}[tb]
\centering
\includegraphics[width=8.5  cm]{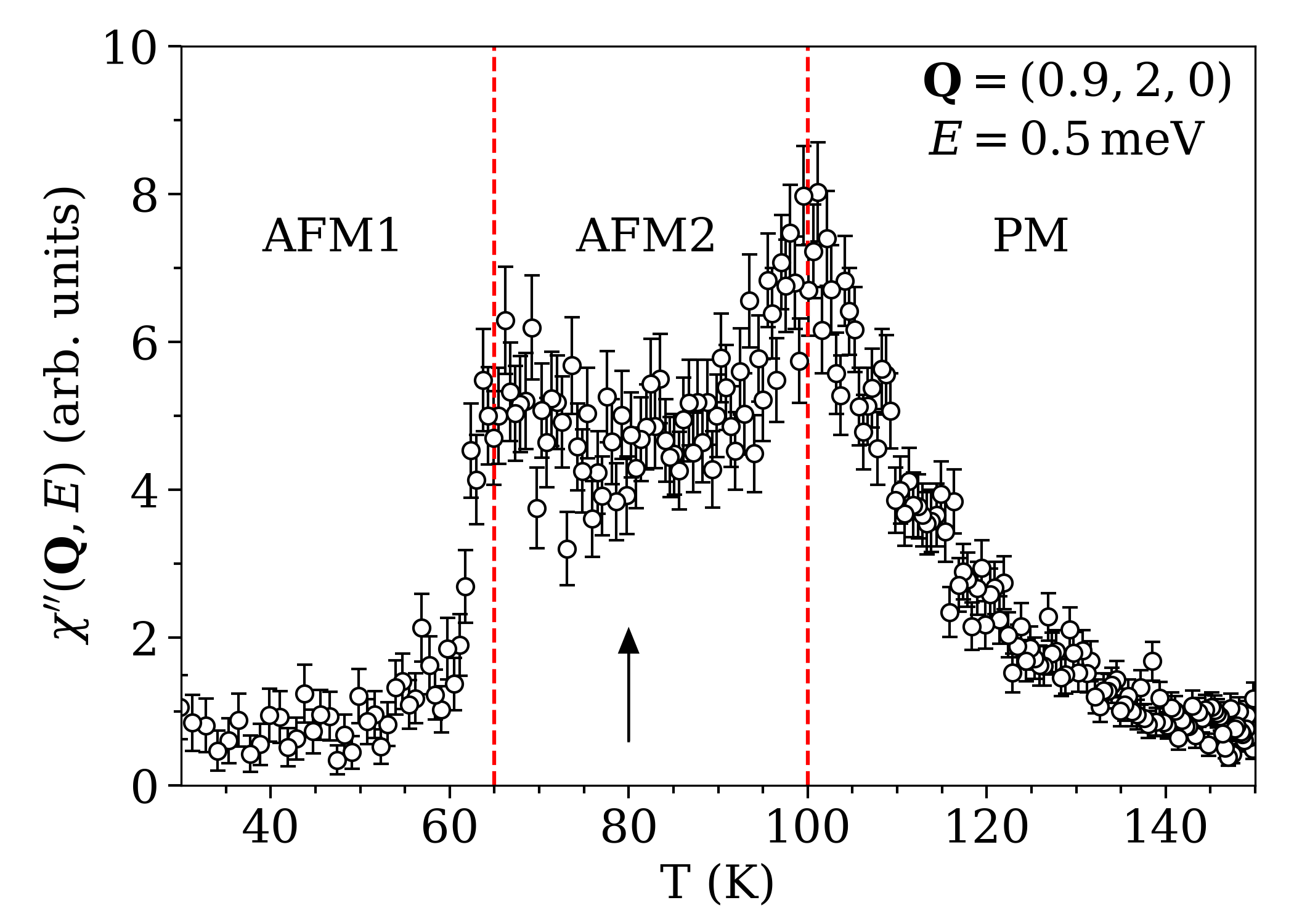}
\caption{Temperature dependence of the dynamical spin susceptibility $\chi''(\VEC Q,E)$ of $\textnormal{Mn}_5\textnormal{Si}_3$ at $\VEC Q = (0.9, 2, 0)$ and $E = 0.5$\,meV measured with unpolarized neutrons with $k_f = 1.5$\,{\AA}$^{-1}$. The vertical red dashed lines indicate $T_{N_{2}}\approx$ 100\,K and $T_{N_{1}}\approx$ 66\,K. The arrow indicates the temperature ($T = 80$\,K) where INS data were collected.}
\label{fig:temp_suscept}
\end{figure}

   To determine the extent of the critical spin fluctuations related to the two AFM transitions, spectra were collected with an unpolarized neutron beam at small $\VEC q$. A ($\VEC Q$, $E$) position was carefully selected to avoid contributions to the measured intensity from the elastic and inelastic scattering from the magnetic zone centers and low energy magnon modes, respectively. The obtained intensity after background subtraction was corrected by the detailed balance factor so that the final result relates to the imaginary part of the dynamical spin susceptibility $\chi''(\VEC Q,E)$. Fig.~\ref{fig:temp_suscept} shows the extent of the spin fluctuations for $\VEC Q = (0.9, 2, 0)$ and $E = 0.5$\,meV in the PM state, as well as, the critical fluctuations due to the two AFM phase transitions as seen by the broad tail above $T_N$. $\chi''(\VEC Q,E)$ shows two maxima at $T_{N_{2}}$ and $T_{N_{1}}$, in agreement with the established magnetic phase diagrams of $\textnormal{Mn}_5\textnormal{Si}_3$ where the AFM transitions occur~\cite{surgers_switching_2017,das_observation_2019}. 

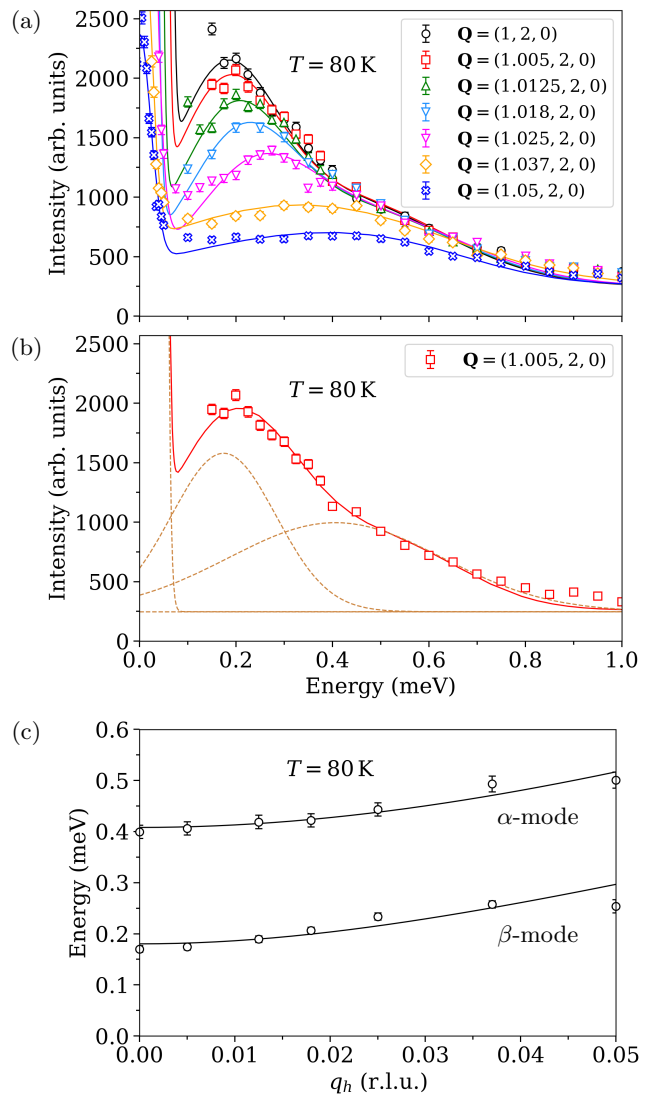
\begin{figure}[tb]
  \setlength{\unitlength}{1cm}
  \newcommand{\boxsize}{0.3}
  \begin{picture}(8,14.65)
    \put( 0.0, 9.57){ \includegraphics[width=8.0cm, trim={0cm 0cm 0cm 0.1cm},clip=true]{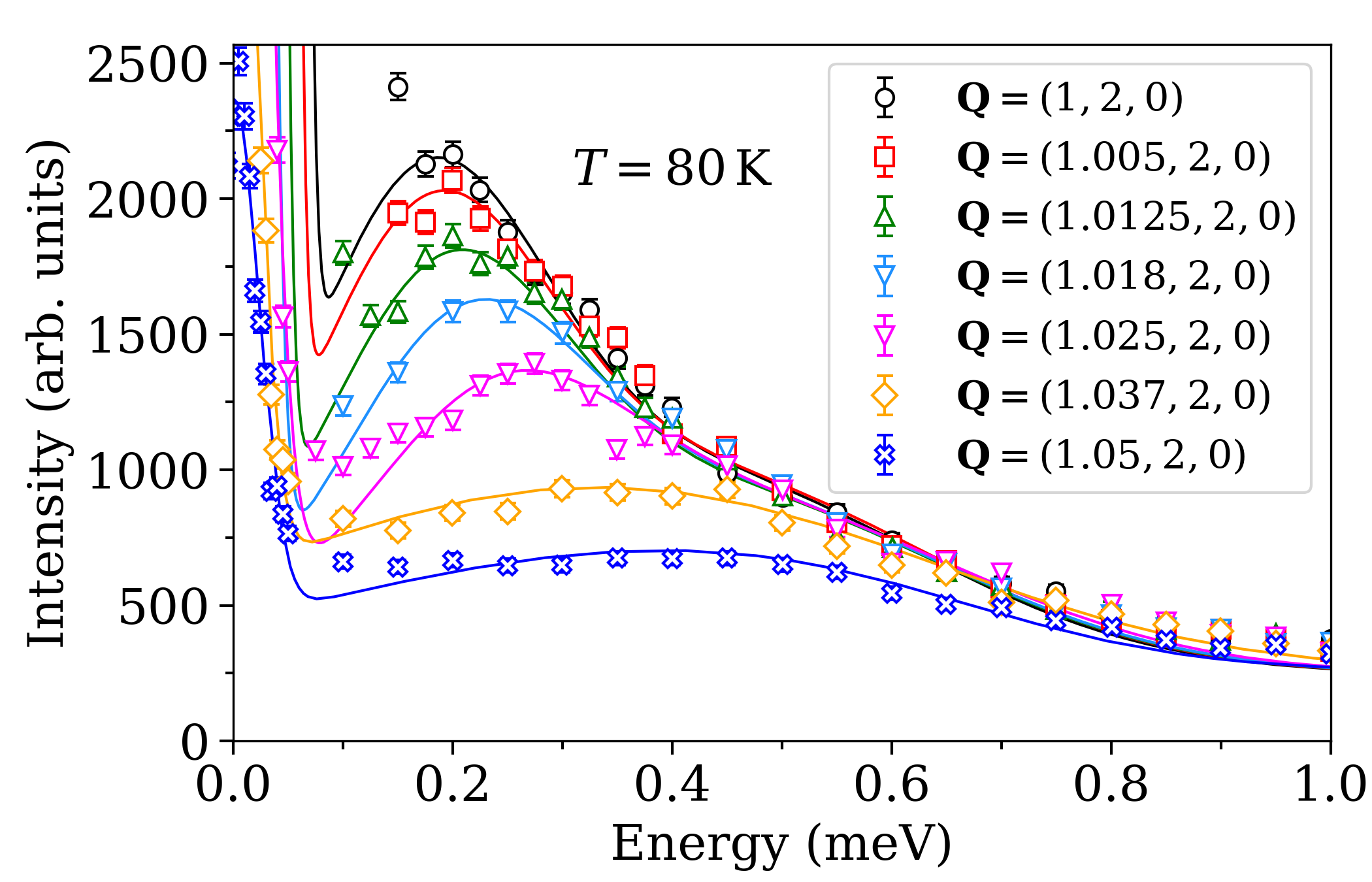} }
    \put( 0.0, 5.25){ \includegraphics[width=8.0cm, trim={0cm 0cm 0cm 0.3cm},clip=true]{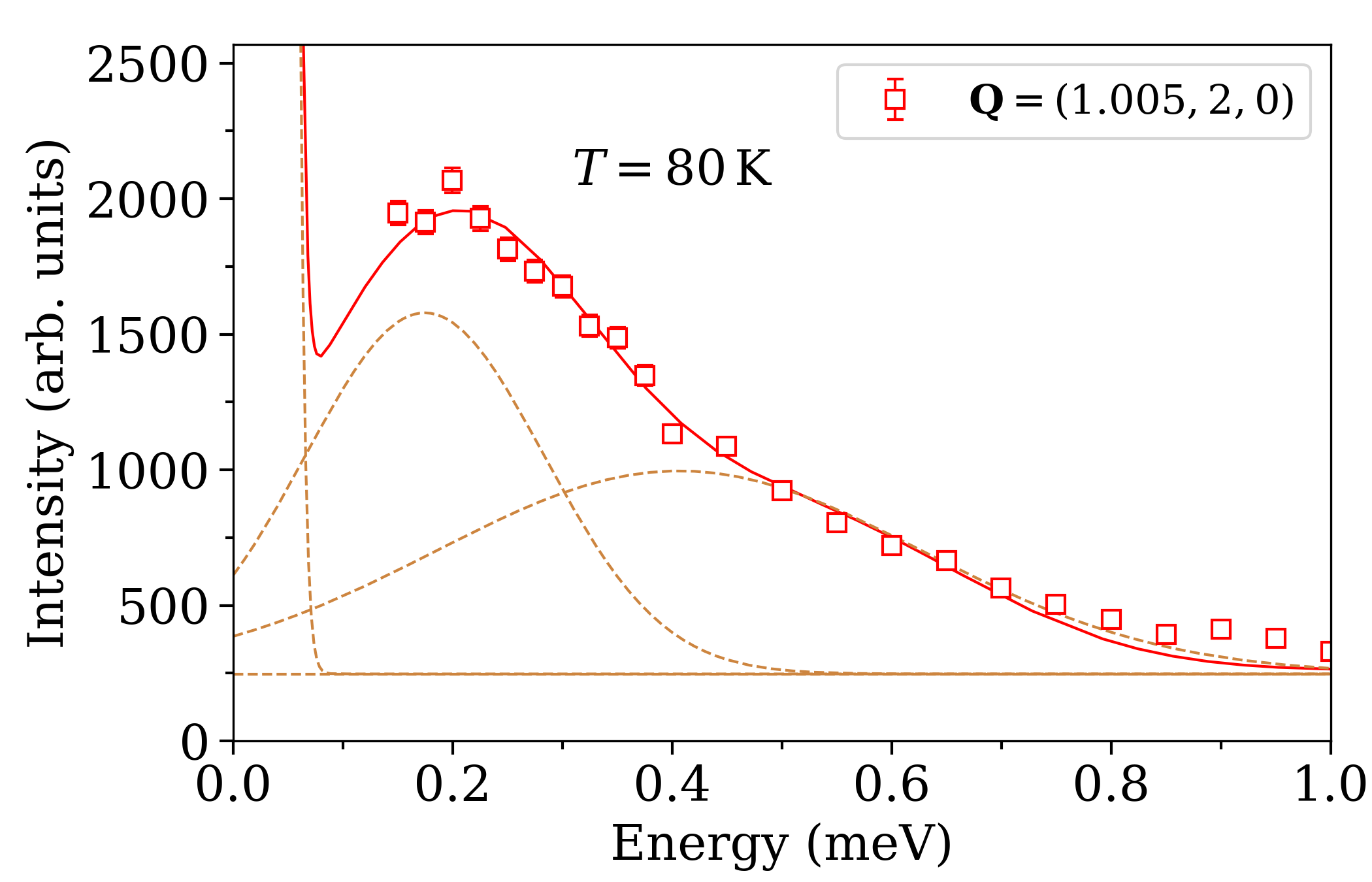} }
    \put( 0.0, 0.00){ \includegraphics[width=8.0cm, trim={0cm 0cm 0cm 0.1cm},clip=true]{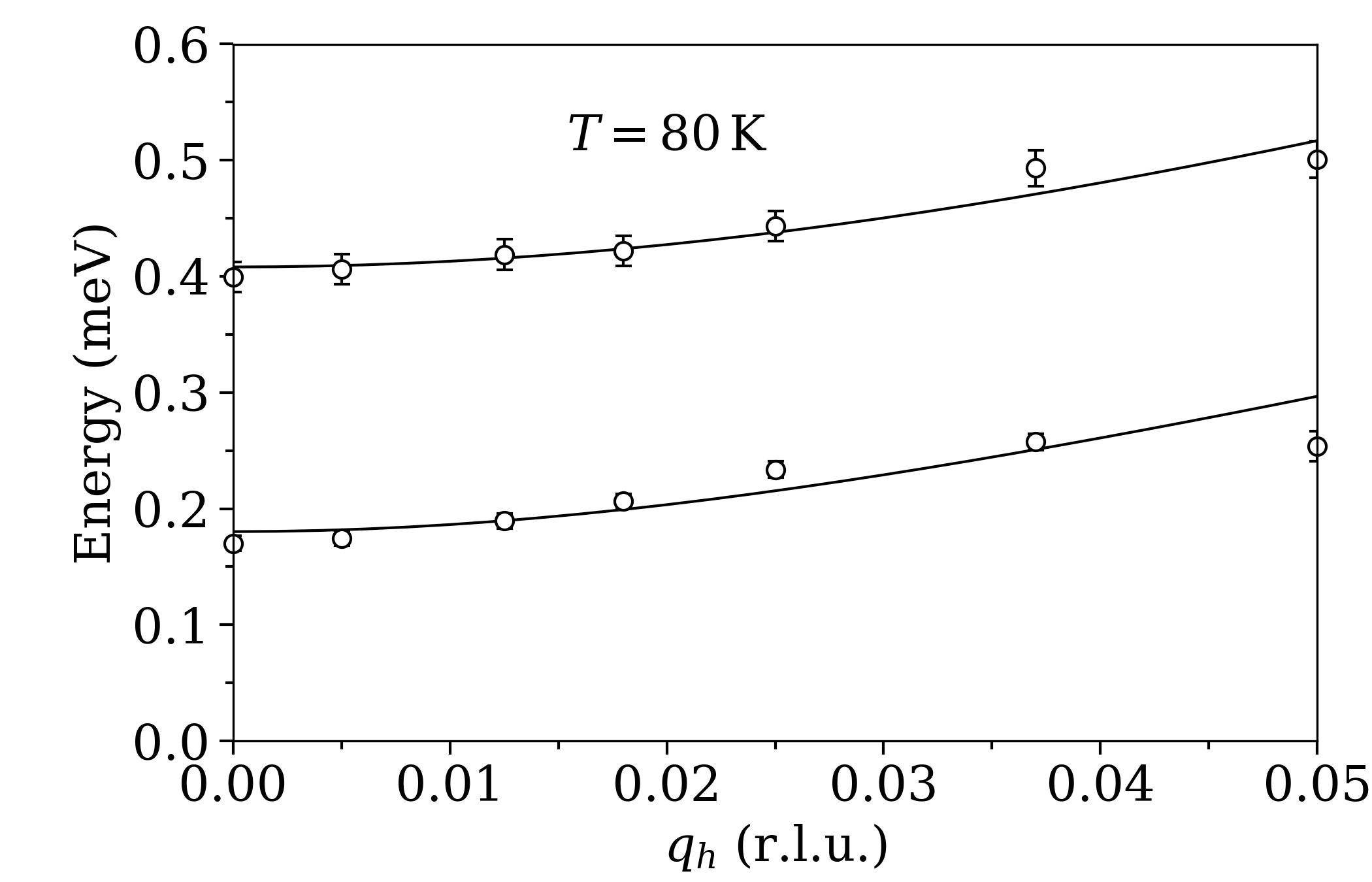} }
    
    \put(6.5, 3.6){ \makebox(\boxsize,\boxsize){$\alpha$-mode} }
    \put(6.5, 2.0){ \makebox(\boxsize,\boxsize){$\beta$-mode} }
    
    \put(-0.3, 14.15){ \makebox(\boxsize,\boxsize){(a)} }
    \put(-0.3,  9.8){ \makebox(\boxsize,\boxsize){(b)} }
    \put(-0.3,  4.7){ \makebox(\boxsize,\boxsize){(c)} }

  \end{picture}
\caption{(a) Energy spectra of Mn$_{5}$Si$_{3}$ at $\VEC Q = (1+q_{h}, 2, 0)$ measured at IN12 spectrometer with unpolarized setup with $k_f = 1.05$\,{\AA}$^{-1}$ at $T = 80$\,K.
(b) Fitting of the $E$--scan at $\VEC Q = (1.005, 2, 0)$. 
The three dashed lines correspond to Gaussian functions sitting on top of a flat background.
The solid lines indicate the overall fits as described in the text. (c) Low energy magnon dispersion at $T = 80$\,K along the ($h$00) direction. The solid lines are fits with the empirical dispersion relation $E=\sqrt{\Delta^2+C^2q^2}$.
}
\label{fig:Capture3}
\end{figure}

   Magnetic excitations were measured around the magnetic zone center $\VEC G = (1, 2, 0)$ at $T = 80$\,K, a temperature selected well inside the AFM2 phase where the intensity shows a plateau and the combined critical spin fluctuations from the PM to the AFM2 ($T_{N_{2}}\approx$ 100\,K) and from the AFM2 to the AFM1 ($T_{N_{1}}\approx$ 66\,K) transitions have minimal intensities (see Fig.~\ref{fig:temp_suscept}). The energy dependence of the measured excitations at different $Q_{h}$ positions is shown in Fig.~\ref{fig:Capture3}(a), where $\VEC Q = (Q_h, 2, 0)$. The individual spectra consist of three peaks. The first peak that is always centered at $E = 0$\,meV corresponds to the elastic line. The two other peaks appear at finite $E$ and shift to higher energy transfers as $Q_{h}$ increases, characteristic of dispersive spin waves. To analyze the obtained spectra, a constant background was assumed and Gaussian functions were selected to describe the peaks. One could argue that the signal at finite $E$ at $\VEC Q  = (1.037, 2, 0)$ and $\VEC Q  = (1.05, 2, 0)$ could be described by a single broad peak. However, the use of polarized neutrons (see Section II.D) justifies the existence of two peaks for the excitations, since the polarized INS cross--sections implore strict fitting conditions. A typical $E$--scan collected at $\VEC Q  = (1.005, 2, 0)$ with the individual fit for the elastic and spin-wave signals is shown in Fig.~\ref{fig:Capture3}(b).

   The obtained low energy experimental spin--wave dispersion along the ($h$00) symmetry direction in the AFM2 phase of Mn$_{5}$Si$_{3}$ is shown in Fig.~\ref{fig:Capture3}(c). There are two characteristic features: (i) two small energy gaps at $\VEC q = 0$ and (ii) the lowest magnetic excitations can be described by the empirical dispersion relation $E=\sqrt{\Delta^2+C^2q^2}$~\cite{Ibuka}, where $\Delta$ refers to the spin--gap and $C$ is a constant. The obtained values for the two magnon modes are: $\Delta_\alpha = 0.408(7)$\,meV, $C_\alpha = 6.3(5)$\,meV/r.l.u., $\Delta_\beta = 0.181(4)$\,meV and $C_\beta = 4.7(3)$\,meV/r.l.u.. The observed gaps are indications of two easy--axis anisotropies, an assumption stemming from the collinear spin arrangement in the AFM2 phase of $\textnormal{Mn}_5\textnormal{Si}_3$. In order to describe the magnon spectrum and to extract the dominant magnetic exchange interactions and magnetocrystalline anisotropy in the AFM2 phase, theoretical calculations were employed (see Section III). In what follows, the modes originating from $\Delta_\alpha$ and $\Delta_\beta$ will be referred to as the $\alpha$ and the $\beta$-modes, respectively.

\subsection{Unpolarized INS measurements under magnetic field}

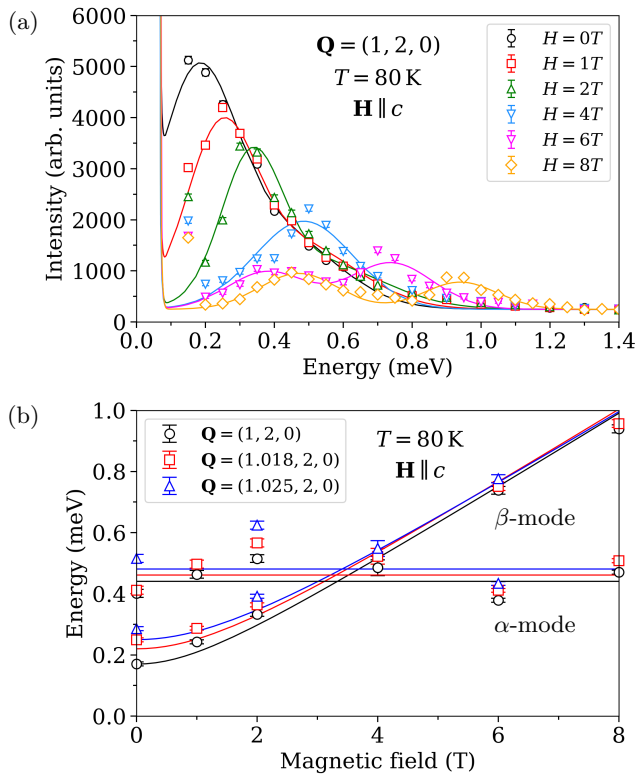
\begin{figure}[tb]
  \setlength{\unitlength}{1cm}
  \newcommand{\boxsize}{0.3}
  \begin{picture}(8,10.35)

    \put( 0.0, 5.25){ \includegraphics[width=8.0cm, trim={0cm 0cm 0cm 0.3cm},clip=true]{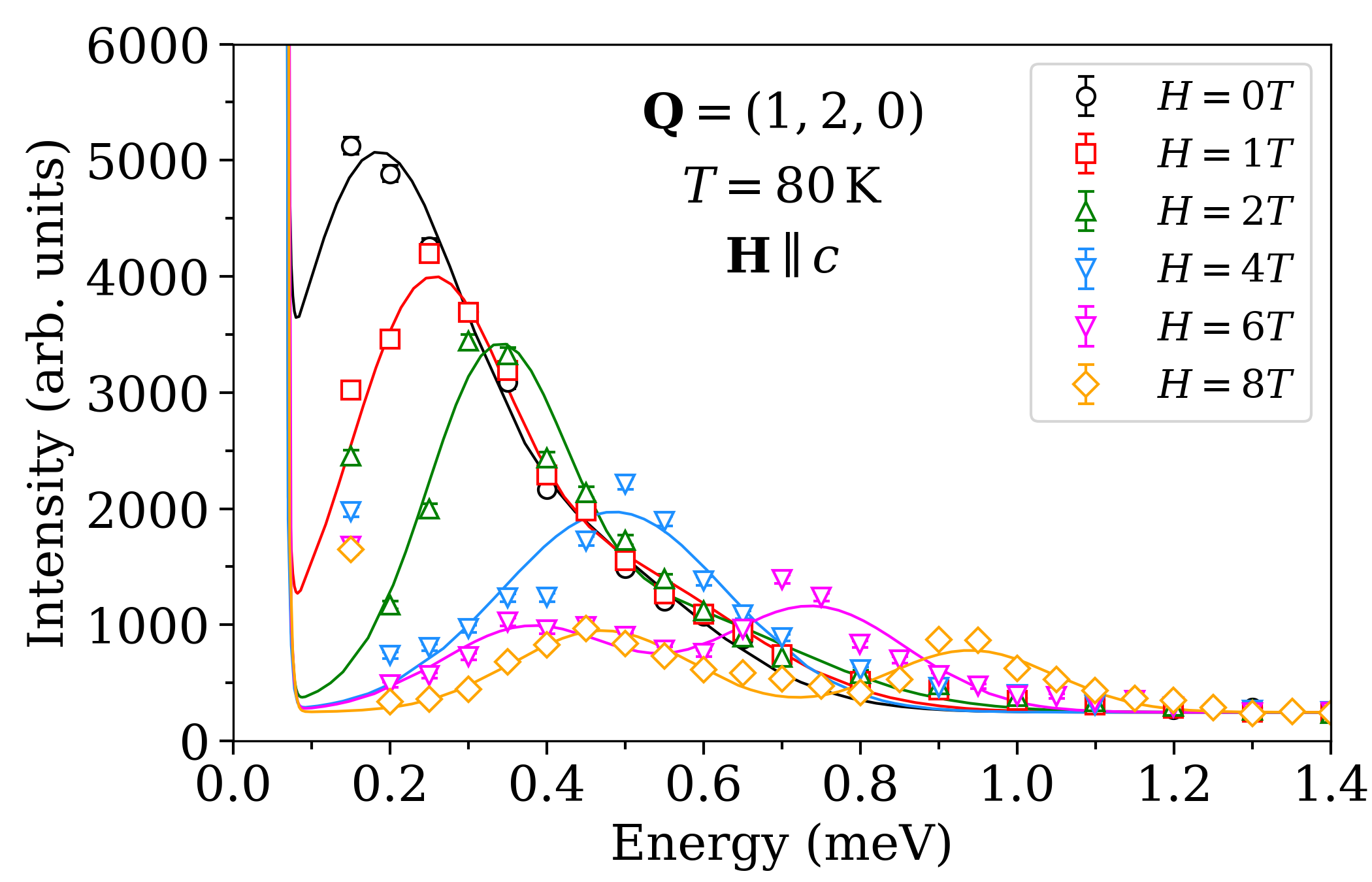} }
    \put( 0.0, 0.00){ \includegraphics[width=8.0cm, trim={0cm 0cm 0cm 0.1cm},clip=true]{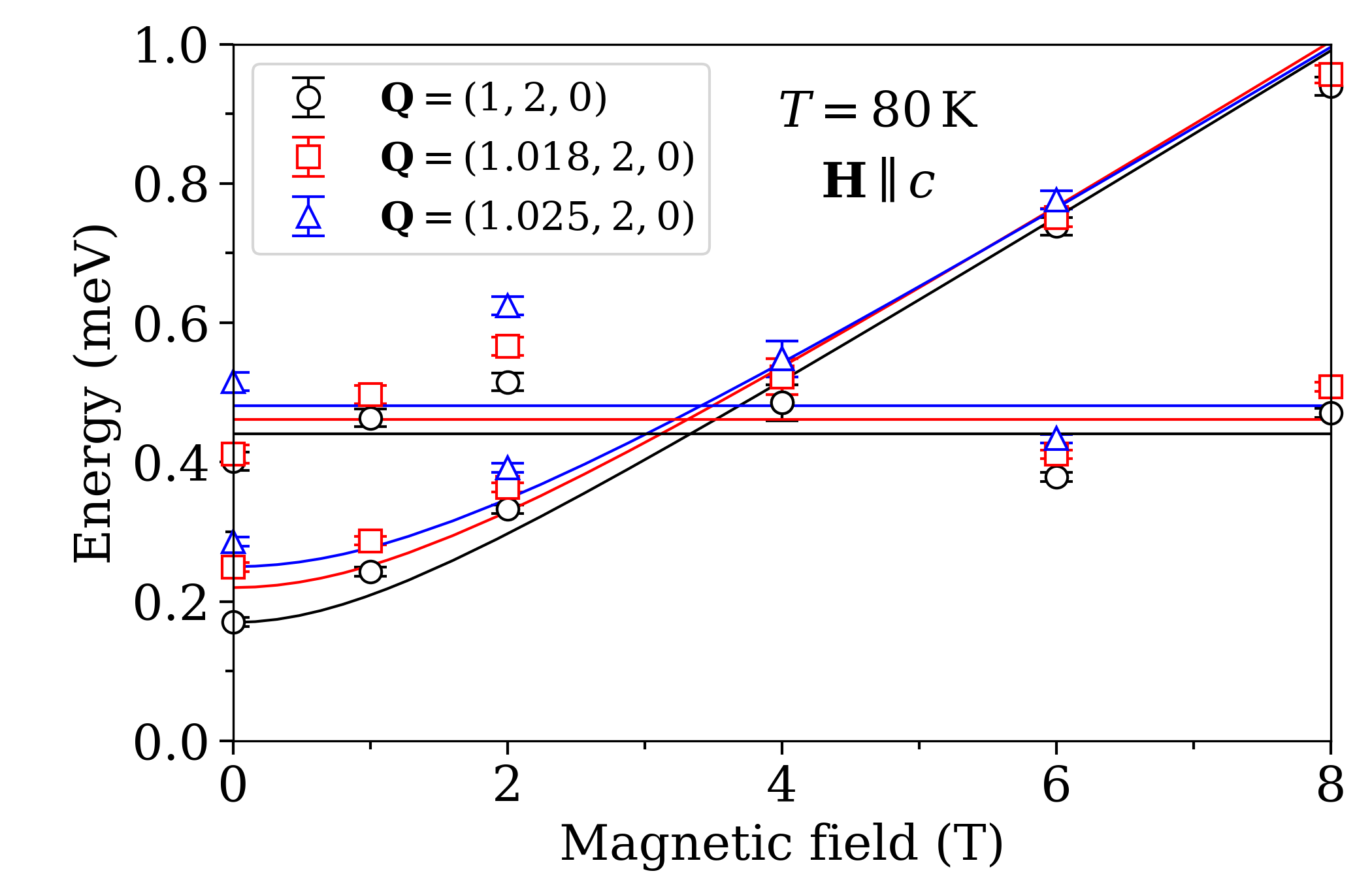} }
    
    \put(6.5, 3.3){ \makebox(\boxsize,\boxsize){$\beta$-mode} }
    \put(6.5, 1.9){ \makebox(\boxsize,\boxsize){$\alpha$-mode} }
    
    \put(-0.3,  9.9){ \makebox(\boxsize,\boxsize){(a)} }
    \put(-0.3,  4.65){ \makebox(\boxsize,\boxsize){(b\textbf{})} }

  \end{picture}
\caption{(a) Inelastic spectra of $\textnormal{Mn}_5\textnormal{Si}_3$ for $\VEC H\parallel \hat{\VEC c}$ obtained at $\VEC Q = (1, 2, 0)$ at $T = 80$\,K (AFM2 phase) with unpolarized beam with $k_f = 1.05$\,{\AA}$^{-1}$ at IN12. The lines indicate fits with Gaussian functions. (b) Energy of the spin excitations as a function of the external magnetic field at three different $Q_{h}$ positions at $T = 80$\,K (AFM2 phase). Lines are guides for the eyes.}
\label{fig:exp_ine_spectra}
\end{figure}

   For $\VEC H\parallel \hat{\VEC c}$, neutron diffraction measurements~\cite{silva_magnetic_2002} performed in single crystals of $\textnormal{Mn}_5\textnormal{Si}_3$ indicate that no field--induced transition occurs within the AFM2 phase and consistently with the macroscopic data~\cite{surgers_switching_2017,luccas_magnetic_2019,das_observation_2019} the PM state is not reached up to $H = 8$\,T due to the steep $T_{N_{2}}(H)$ phase boundary. In order to investigate the magnon spectrum under magnetic field, energy spectra were collected at three different $Q_{h}$ positions (1, 1.018 and 1.025\,r.l.u.) around the magnetic Bragg peak $\VEC G = (1, 2, 0)$. Figure~\ref{fig:exp_ine_spectra}(a) shows such characteristic scans at $Q_{h} = 1$\,r.l.u. for different magnetic fields. The obtained spectra were analyzed as described in the previous section. For increasing magnetic field, the two peaks that correspond to the $\alpha$ and the $\beta$--modes show different behavior. As the field increases, the position and intensity of the $\alpha$--mode is not significantly affected in contrast to the $\beta$--mode, which disperses with the field and a continuous diminution of intensity is observed. It should be noted that at $H = 4$\,T, the two peaks seem to merge. 

   The same observations regarding the behavior of the two modes stem from measurements at $Q_{h} = 1.018$\,r.l.u. and $Q_{h} = 1.025$\,r.l.u. and are summarized in Fig.~\ref{fig:exp_ine_spectra}(b). The obtained results indicate that the modes behave differently under the external magnetic field possibly due to their different polarization. To shed light on this behavior, spectra were collected using the polarized INS method.

\subsection{Polarized INS measurements}

   As a general rule, neutron scattering is only sensitive to magnetic excitations perpendicular to $\VEC Q$~\cite{chatterji_neutron_2006}. With LPA it is possible to separate magnetic fluctuations polarized along different directions in spin space. The initial polarization was prepared parallel to $x$--axis, perpendicular to $\VEC Q$ in the scattering plane ($y$--axis) and perpendicular to the scattering plane ($z$--axis), and the final polarization was analyzed for a scattering process reversing the initial polarization by 180\,$^{\circ}$. The corresponding measurement channels are canonically labeled SF$_{xx}$, SF$_{yy}$, and SF$_{zz}$, where SF stands for ``Spin--Flip''. 

   The neutron scattering double differential cross--sections for the three SF channels are~\cite{chatterji_neutron_2006}:
\begin{eqnarray}
\text{SF}_{xx}=\left(\frac{d^2\sigma}{d\Omega dE}\right)_{\text{SF}}^{x} \propto \text{BG}_{\text{SF}} + \langle{\delta}{M}_{y}\rangle + \langle{\delta}{M}_{z}\rangle\\
\text{SF}_{yy}=\left(\frac{d^2\sigma}{d\Omega dE}\right)_{\text{SF}}^{y} \propto \text{BG}_{\text{SF}} + \langle{\delta}{M}_{z}\rangle\\
\text{SF}_{zz}=\left(\frac{d^2\sigma}{d\Omega dE}\right)_{\text{SF}}^{z} \propto \text{BG}_{\text{SF}} + \langle{\delta}{M}_{y}\rangle
\end{eqnarray}
where $\text{BG}_{\text{SF}}$ is the background (which includes the nuclear spin scattering) and $\langle{\delta}{M}_{y}\rangle$ and $\langle{\delta}{M}_{z}\rangle$ the measured magnetic fluctuations. Considering that: (i) $\VEC Q$ is in the $ab$--plane, (ii) in the AFM2 phase the magnetic moments lie parallel and antiparallel to the $b$--axis and (iii) spin--waves correspond to precession perpendicular to the ordered moment, then in the crystal frame the cross--sections become:
\begin{eqnarray}
\text{SF}_{xx} \propto \text{BG}_{\text{SF}} +\sin^2\theta\langle{\delta}{M}_{a}\rangle + \langle{\delta}{M}_{c}\rangle\\
\text{SF}_{yy} \propto \text{BG}_{\text{SF}} + \langle{\delta}{M}_{c}\rangle\\
\text{SF}_{zz} \propto \text{BG}_{\text{SF}} +\sin^2\theta\langle{\delta}{M}_{a}\rangle
\end{eqnarray}
with $\theta$ the angle between $\VEC Q$ and the [100] direction and can be calculated by $\theta = \text{arctan}(\frac{Q_{k}}{Q_{h}}\frac{a}{b}$).

\begin{figure}[tb]
  \setlength{\unitlength}{1cm}
  \newcommand{\boxsize}{0.3}
  \begin{picture}(8,5.1)
    \put(0, 0){\includegraphics[width=8cm]{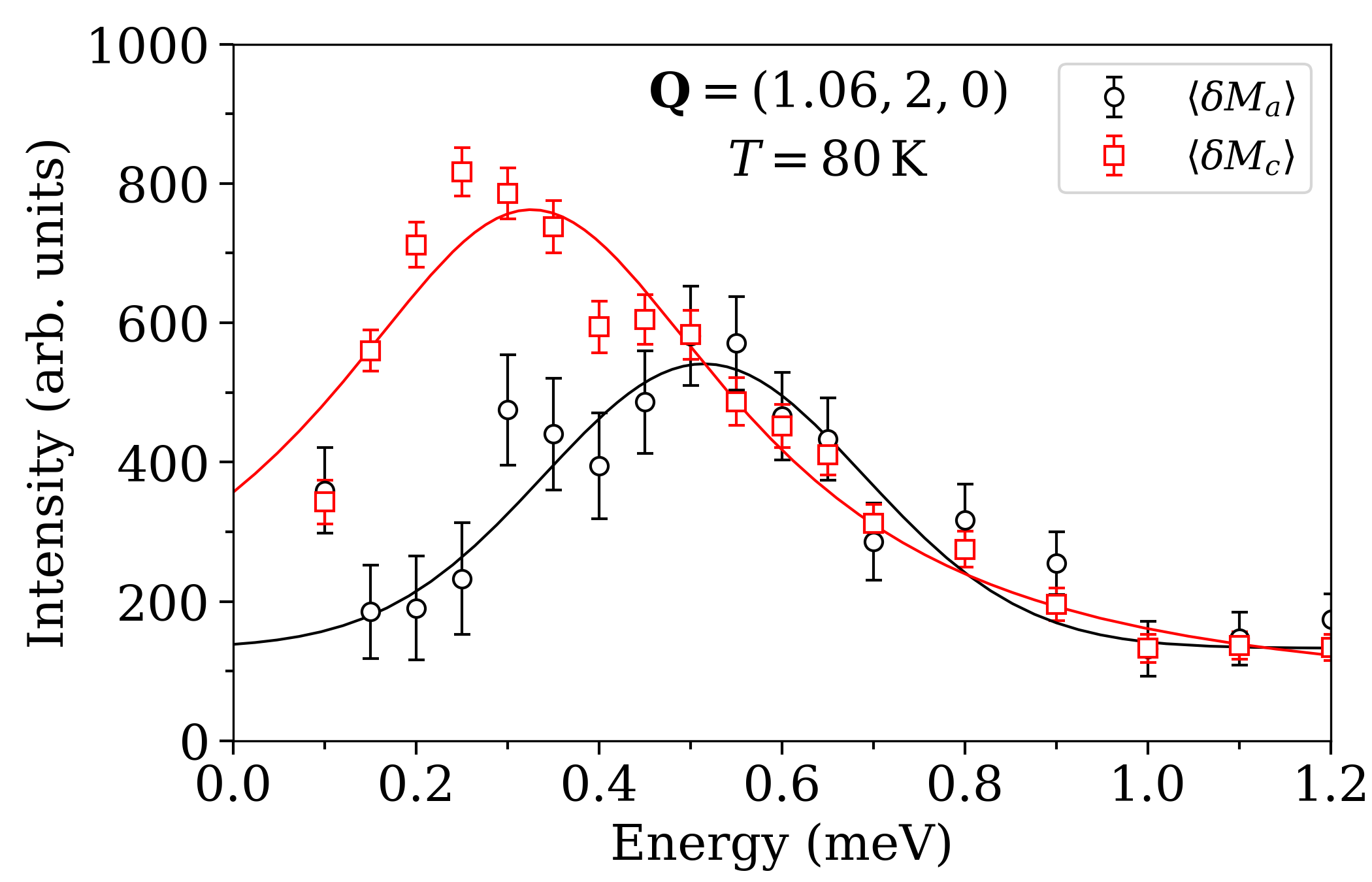} }

    \put(2.6, 4.35){ \makebox(\boxsize,\boxsize){$\beta$-mode} }
    \put(4.05, 3.55){ \makebox(\boxsize,\boxsize){$\alpha$-mode} }
  \end{picture}
\caption{Subtracted spin fluctuations spectra $\langle{\delta}{M}_{a}\rangle$ and $\langle{\delta}{M}_{c}\rangle$ of $\textnormal{Mn}_5\textnormal{Si}_3$ obtained at ThALES and measured at $\VEC Q = (1.06, 2, 0)$ at $T = 80$\,K. The intensity for $\langle{\delta}{M}_{a}\rangle$ was corrected by the factor $\text{sin}^2\theta = 0.545$. The lines indicate fits with Gaussian functions.}
\label{fig:Capture5}
\end{figure}

   To get further insight regarding the polarization dependence of the two magnon modes, $E$--spectra at different $\VEC Q$ positions were collected in the three SF channels at $T = 80$\,K. The magnetic fluctuations $\langle{\delta}{M}_{a}\rangle$ and $\langle{\delta}{M}_{c}\rangle$ were extracted by taking the difference of intensities between the different polarization channels. A typical result of such analysis is shown in Fig.~\ref{fig:Capture5} at $\VEC Q = (1.06, 2, 0)$ where the intensity for $\langle{\delta}{M}_{a}\rangle$ is corrected by the angle prefactor. The peak positions of the subtracted spin fluctuations spectra are consistent with the established low energy magnon dispersion curves obtained from the unpolarized data and shown in Fig.~\ref{fig:Capture3}(c). Moreover, it is evident that the maximum of intensity in $\langle{\delta}{M}_{a}\rangle$ and in $\langle{\delta}{M}_{c}\rangle$ corresponds to the $\alpha$--mode and $\beta$--mode, respectively. This hints that the elliptic polarization of each mode is different and points along different crystal axis.  
   
\section{Theoretical calculations}

\subsection{Density functional theory}
First--principles calculations were performed to determine the ground--state electronic and magnetic properties of the AFM2 phase of $\textnormal{Mn}_5\textnormal{Si}_3$. 
Our study was based on the atomic structure specified in Ref.~\onlinecite{brown_antiferromagnetism_1995}.
We employed density functional theory (DFT) using the full--potential Korringa--Kohn--Rostoker Green--function (KKR--GF) method including spin--orbit coupling, as implemented in the \textit{JuKKR} code~\cite{papanikolaou_conceptual_2002}, using the local--spin---density approximation~\cite{Vosko1980}.
The cutoff for the angular momentum expansion of the scattering problem was set to $l_\textnormal{max} = 3$.
Furthermore, the energy integration was performed in the upper complex energy plane~\cite{wildberger_fermi-dirac_1995} with 30 points in a rectangular path and 5 Matsubara frequencies at a temperature of $T=\SI{473.68}{\K}$, and the Brillouin zone integration was performed with $30 \times 15 \times 30$ $k$--points.
The magnetic exchange tensor, which parametrizes the spin Hamiltonian discussed in the following section, was obtained through the \textit{infinitesimal--rotations} method~\cite{liechtenstein_local_1987,Ebert2009}.
In these calculations, the number of Matsubara frequencies was increased to 10 with $T=\SI{100}{\K}$.

We explored two possible magnetic configurations with self--consistent calculations, the AFM configuration that is shown in Fig.~\ref{fig:Capture1}, and a ferromagnetic (FM) phase with finite magnetic moments in both Mn1 and Mn2 sites.
The AFM was found the most energetically favorable with the FM phase being 146\,meV per unit cell higher in energy.
In the FM phase, the magnetic moments are 2.6 and 1.2\,$\muB$ for the Mn2 and Mn1 sites, respectively, while the Si sites have magnetic moments of 0.1\,$\muB$ antiparallel to those of the manganese sites.
For the AFM phase, two--thirds of the Mn2 site carry magnetic moments of 2.4\,$\muB$ while the other Mn and the Si atoms have no magnetic moment.

Combining the magnetic force theorem with the frozen potential approximation, the magnetocrystalline anisotropy was determined from band energy differences between states with different orientations of the spin magnetic moments.
For the AFM phase, we obtained the following energy differences when aligning the magnetic moments along the main crystal axes: $ E^a - E^b = 0.12$, $ E^c - E^b = 0.09$, and $E^a - E^c = \SI{0.03}{\milli\electronvolt}$ per magnetic atom, which indicates that $b$ and $c$ are the first and second preferred axis, respectively, and $a$ is the hard--axis.

\subsection{Model Hamiltonian and spin--wave calculation}

We mapped the \textit{ab initio} calculations onto a quantum Heisenberg Hamiltonian to study the spin--wave excitations in the adiabatic approximation, which reads as
\begin{equation}\label{eq:heisenberg_hamiltonian}
    \HH = - \sum_{ij} J_{ij} \VEC S_i \cdot \VEC S_j  - \sum_\alpha k^\alpha \sum_i (S^\alpha_i)^2 .
\end{equation}
The first term is due to the magnetic exchange interaction whose coupling is given by $J_{ij}$.
$\VEC S_i$ is the spin for which we set $S = 1$.
The second term accounts for the biaxial magnetocrystalline anisotropy ($\HH_\textnormal{ani}$) with $k^b = E^a - E^b = \SI{0.12}{\milli\electronvolt}$ and $k^c = E^a - E^c = \SI{0.03}{\milli\electronvolt}$, both being positive.
From the DFT calculations, we obtained that $k^b > k^c$, which makes $b$ the primary easy--axis, and $c$ the secondary easy--axis.
The biaxial anisotropy could also be modeled by a combination of an easy--axis along $b$ and an easy--plane ($bc$--plane) anisotropy.

The magnetic exchange interactions were obtained from first--principles calculations for the AFM2 phase.
The results for the first few Mn2 pairs as indicated in Fig.~\ref{fig:Capture1}(a) are:
$J_1 = -12.23$, $J_2 = -2.16$, $J_3 = 3.98$ and $J_4 = \SI{-2.89}{\meV}$. 
Most interactions, apart from $J_3$, have AFM character.
This indicates that the AFM2 phase is favored by all those pair interactions, that is, within this set of interactions, there is no frustration. $J_1$, $J_2$, and $J_3$ correspond to couplings between magnetic moments in the same [Mn2]${_6}$ octahedra.
$J_1$ has the highest value and is the exchange interaction between the spins located on a triangle in the $ab$--plane (distance 2.789\,\AA).
$J_2$ and $J_3$ couple spins located on adjacent triangles separated by $c$/2 with distances 2.893 and 4.019\,\AA, respectively. The exchange interaction $J_4$ concerns the shortest distance (4.364\,\AA) between spins located on adjacent [Mn2]${_6}$ octahedra.

We employ the linear spin--wave approximation to obtain the spin--wave excitations of the quantum Heisenberg Hamiltonian using the computed magnetic interaction parameters.
The spin--wave excitations are the eigenstates of the dynamical matrix associated with the quantum Heisenberg Hamiltonian in Eq.~\eqref{eq:heisenberg_hamiltonian}, as explained in detail in Ref.~\onlinecite{dos_santos_spin-resolved_2018}.
We start by constructing a local coordinate system for every magnetic site with the local $z$--axis coinciding with the classical ground--state spin orientation.
In this local frame, we expand the quantum spin operators using the linearized Holstein--Primakoff transformation as $\VEC S_i = \left(\sqrt{2S}\frac{a_i+a^\dagger_i}{2}, \sqrt{2S}\frac{a_i-a^\dagger_i}{2\mathrm{i}}, S - a^\dagger_i a_i \right)$, where $a^\dagger_i$ and $a_i$ are bosonic ladder operators~\cite{holstein_field_1940}.
Keeping only terms up to second order in the Holstein--Primakoff bosons, the Hamiltonian can be written as $\MC H = \MC H_0 + \MC H_2$.
The $\MC H_0$ term is a constant corresponding to the classical ground--state energy.
The second term
\begin{equation}
    \HH_2 = - \sum_\VEC k \sum_{\mu\nu} \VEC a ^\dagger (\VEC k) \VEC H_{\mu\nu}(\VEC k)\VEC a_\nu(\VEC k)
\end{equation}
contains the quadratic terms of the Holstein--Primakoff bosons describing the spin excitations, where $\VEC H (\VEC k)$ is a $2n \times 2n$ matrix and $\VEC a_\mu(\VEC k) = \icol{a_\mu(\VEC k) \\  a^\dagger_\mu(-\VEC k)} $ with $a_\mu(\VEC k) = \frac{1}{\sqrt{N}} \sum_m e^{-\iu \VEC k \cdot \VEC R_m} a_{m\mu}$.
$n$ and $N$ are the number of atoms in the unit and the number of unit cells under periodic Born--von--Karman boundary conditions, respectively.
The spin--wave eigenvalues $\omega(\VEC k)$ and eigenvectors $\ket{\VEC k}$ are then obtained by a Bogoliubov transformation~\cite{bogoljubov_new_1958}.
This process diagonalizes the system's dynamical matrix $\VEC D = \VEC g \HH_2$, where $\VEC g$ is a diagonal matrix containing $-1$ on its first half and 1 on the second while ensuring the bosonic character of the diagonalizing basis.
The spin--wave inelastic scattering spectrum is computed with our theory for spin--resolved electron--energy--loss spectroscopy (SREELS) of noncollinear magnets presented in Ref.~\onlinecite{dos_santos_spin-resolved_2018}, where we employ time--dependent perturbation theory to describe the interaction between the probing beam and the magnetic excitations.
The same theory can be applied with little modification to describe inelastic neutron scattering.
This method has been applied to investigate ferromagnetic and antiferromagnetic noncollinear spin textures in Refs.~\onlinecite{dos_santos_nonreciprocity_2020,dos_santos_modeling_2020,dos_santos_first-principles_2017}.

\section{Comparing experimental and theoretical results}

\subsection{Results without external magnetic field}

As already mentioned, the DFT calculations found that the collinear AFM2 phase is more stable than the FM one, in line with the experimental finding that the collinear AFM2 phase is stable in an intermediate temperature range~\cite{brown_antiferromagnetism_1995}.
The AFM2 phase has the peculiarity that one--third of the Mn2 sites and all the Mn1 sites have vanishing ordered magnetic moments, with the main question being whether this is due to a collapse of the local magnetic moment or due to a disordering of their orientation.
In the DFT calculations, we considered the collapse scenario, which did not lead to a strong energetic penalty, given that the FM phase was still substantially higher in energy.
However, the obtained strong AFM exchange interactions between the Mn2 moments in the [Mn2]${_6}$ octahedra (the stacked triangles in Fig.~\ref{fig:Capture1}) can also lead to a finite temperature scenario where a third of the Mn2 moments is orientationally disordered, as proposed for Mn$_3$Pt in Ref.~\onlinecite{MendiveTapia2019}.
The DFT calculations can thus also support a hybrid scenario where orientational disorder is mixed with sizable fluctuations of the magnitude of the moments for the Mn2 sites with vanishing ordered moments.
Concerning the Mn1 sites, the DFT calculations suggest that their magnetic moments are unstable on their own, and come into existence depending on the Mn2 environment.
When this environment is FM a sizable magnetic moment arises in the Mn1 sites, while if this environment is AFM the magnetic moment collapses.
We conclude that the DFT calculations do not rule out the existence of strong spin fluctuations either on the Mn1 or on the Mn2 sites.
Previous experimental work~\cite{biniskos_spin_2018} has suggested that the magnetic excitation spectrum of the AFM2 phase consists of propagating spin--waves and diffuse spin fluctuations, originating from the presence of magnetic and nonmagnetic Mn sites within this phase, respectively.
Further investigation is required to enlighten this phenomenon.

\begin{figure}[tb]
  \setlength{\unitlength}{1cm}
  \newcommand{\boxsize}{0.3}
  \begin{picture}(8,6.2)
    \put( 0, 0){ \includegraphics[width=8.0cm, trim={0cm 0cm 0cm 0cm},clip=true]{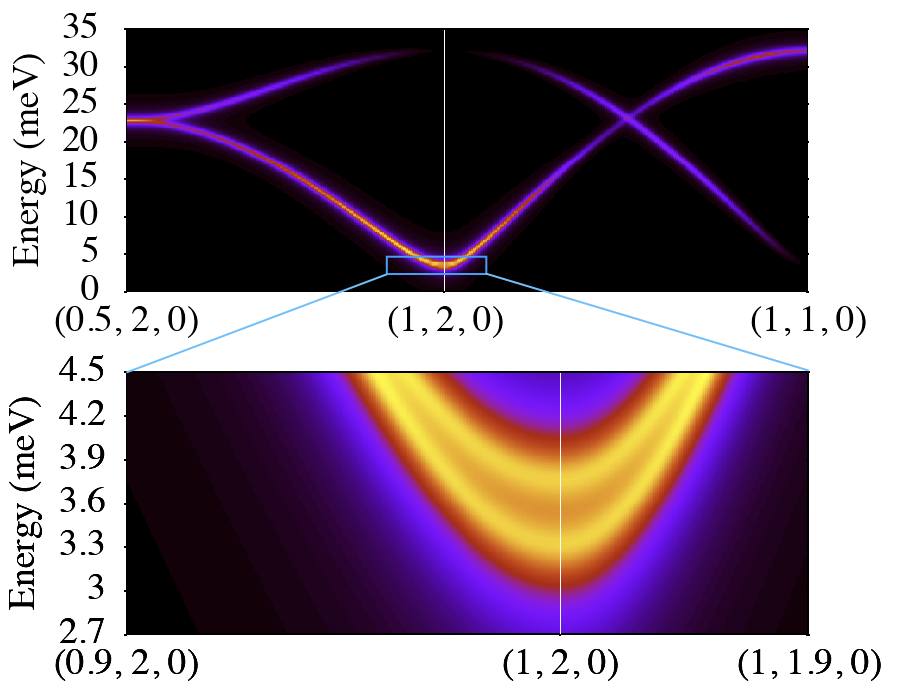} }
    
    \put(-0.3, 5.8 ){ \makebox(\boxsize,\boxsize){(a)} }
    \put(-0.3, 2.8){ \makebox(\boxsize,\boxsize){(b)} }

  \end{picture}
  \caption{\label{fig:theory_spinwave_dispersion}
  (a) Theoretical inelastic scattering spectrum for the AFM2 phase of $\textnormal{Mn}_5\textnormal{Si}_3$ with parameters obtained from DFT calculations in the absence of an external field.
  (b) Zooming in the low energy regions of the spectrum a splitting of the spin--wave modes is observed (double spin--gap) due to the system's biaxial anisotropy.
  }
\end{figure}

The INS data obtained in the collinear AFM2 phase of $\textnormal{Mn}_5\textnormal{Si}_3$ have demonstrated a spin excitation double energy gap for zero field at $\VEC Q = (1,2,0)$, see Fig.~\ref{fig:exp_ine_spectra}(b).
That is, the low energy spectrum is composed of two excitations of nonvanishing and distinct energies. In a system with uniaxial anisotropy, these two excitations would be degenerate.
Using the parameters obtained in our first--principles simulations and given in the previous sections, we calculated the inelastic scattering spectrum for this phase of our material, as seen in  Fig.~\ref{fig:theory_spinwave_dispersion}(a).
In the low energy region around the Brillouin zone center $\VEC G = (1,2,0)$, we observe an energy gap of about 3\,meV.
However, a closer look reveals the double energy--gap structure which is due to the second easy--axis anisotropy predicted in our calculations, as demonstrated by  Fig.~\ref{fig:theory_spinwave_dispersion}(b).

Despite this qualitative agreement, these energy gaps are much higher than those in Fig.~\ref{fig:exp_ine_spectra}(b) obtained through the neutron scattering measurements.
We believe that this discrepancy comes from the difficulties associated with modeling the spin--wave spectrum at finite temperature.
For example, the computed magnetic moments (2.4\,$\muB$) are higher than the ones obtained experimentally (1.48\,$\muB$)~\cite{brown_antiferromagnetism_1995}, which could be explained by the thermal fluctuations of their orientations which have not been theoretically considered.
Possible effects associated with the spin fluctuations of the Mn1 or Mn2 sites are also not taken into account by our model Hamiltonian.
Another source of uncertainty is how temperature may affect the magnetic interactions that define the spin--wave excitation spectrum.
For instance, the effective magnetic anisotropy energy was found to be strongly temperature--dependent in Mn$_5$Ge$_3$~\cite{Maraytta2020}, the ferromagnetic counterpart of Mn$_5$Si$_3$.

To allow a direct comparison between the theoretical spin--wave energies and the experimental data, we rescaled the DFT parameters to match the experimental results. 
A good agreement was obtained by scaling down uniformly the magnetic exchange interaction and the anisotropy parameters by a factor of ten. 
For fine tuning, the second easy--axis anisotropy parameter was adjusted from the scaled DFT value $k^c = \SI{0.003}{\meV}$ to $k^c = \SI{0.009}{\meV}$.

\subsection{Results under external magnetic field}

Using the rescaled parameters, we calculated the spin--wave energies as a function of the applied magnetic field to shed light on the observed behavior of the two low energy magnon modes.
To this aim, we included in our model Hamiltonian the Zeeman term:
\begin{equation}
\HH_\textnormal{Z} = - 2 \muB \sum_i \VEC H \cdot \VEC S_i ,
\end{equation}
to account for an external magnetic field applied along the $c$ and $a$ crystal axes.
It is worth mentioning that previous investigations in AFM systems with uniaxial and biaxial anisotropy have described the influence of an external magnetic field in the magnon properties~\cite{keffer_theory_1952,rezende_introduction_2019}. 
However, the observed behavior from our INS results shown in Fig.~\ref{fig:exp_ine_spectra}(b) is not sufficiently covered.

\begin{figure}[tb]
  \setlength{\unitlength}{1cm}
  \newcommand{\boxsize}{0.3}
  \begin{picture}(8,13.3)
    \put( 0.0, 7.65){ \includegraphics[width=8.0cm, trim={0cm 0cm 0cm 0cm},clip=true]{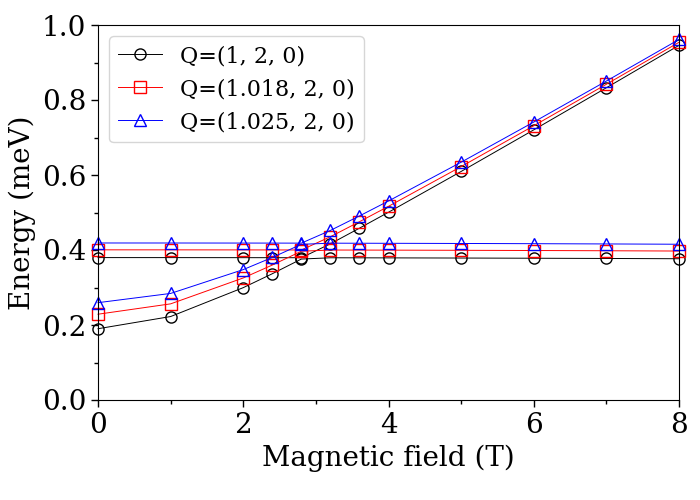} }
    \put( 0.0, 3.00){ \includegraphics[width=8.0cm, trim={0cm 0cm 0cm 0.1cm},clip=true]{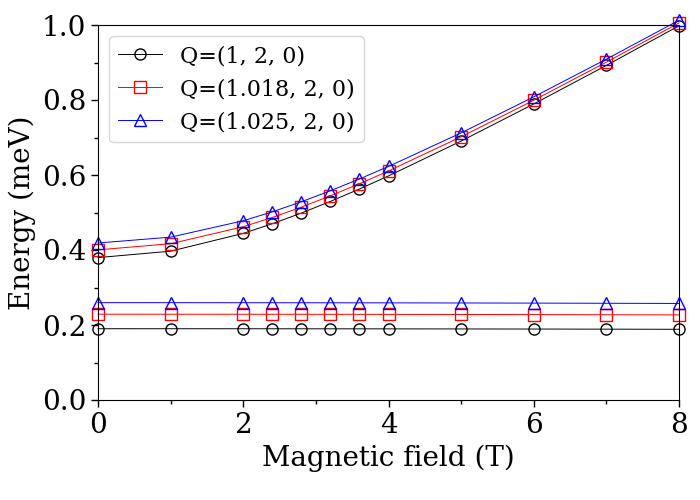} }
    
    \put(   4.6, 12.4 ){\color{black}\scalebox{1.3}{\makebox{$\VEC H \parallel c$}} }
    \put(   4.6, 7.8) {\color{black}\scalebox{1.3}{\makebox{$\VEC H \parallel a$}} }
    \put( 6.5, 9.95){ \makebox(\boxsize,\boxsize){$\alpha$-mode} }
    \put( 6.5, 11.45){ \makebox(\boxsize,\boxsize){$\beta$-mode} }
    \put(6.5, 7.15){ \makebox(\boxsize,\boxsize){$\alpha$-mode} }
    \put(6.5, 5.5){ \makebox(\boxsize,\boxsize){$\beta$-mode} }
    
    \linethickness{0.2cm}
    \put(1.1,8.6){ \color{white} \line(1,0){6.7}}
    
    \put(-0.3, 12.9){ \makebox(\boxsize,\boxsize){(a)} }
    \put(-0.3,  8.2){ \makebox(\boxsize,\boxsize){(b)} }
    \put( 0.5,  2.7){ \makebox(\boxsize,\boxsize){(c)} }
    \put( 4.5,  2.7){ \makebox(\boxsize,\boxsize){(d)} }
    
    \put( 1.0, 0){
        \put( 0.0, 0){\includegraphics[width=2.3cm, trim={2.2cm 0cm -0.4cm 0cm},clip=true]{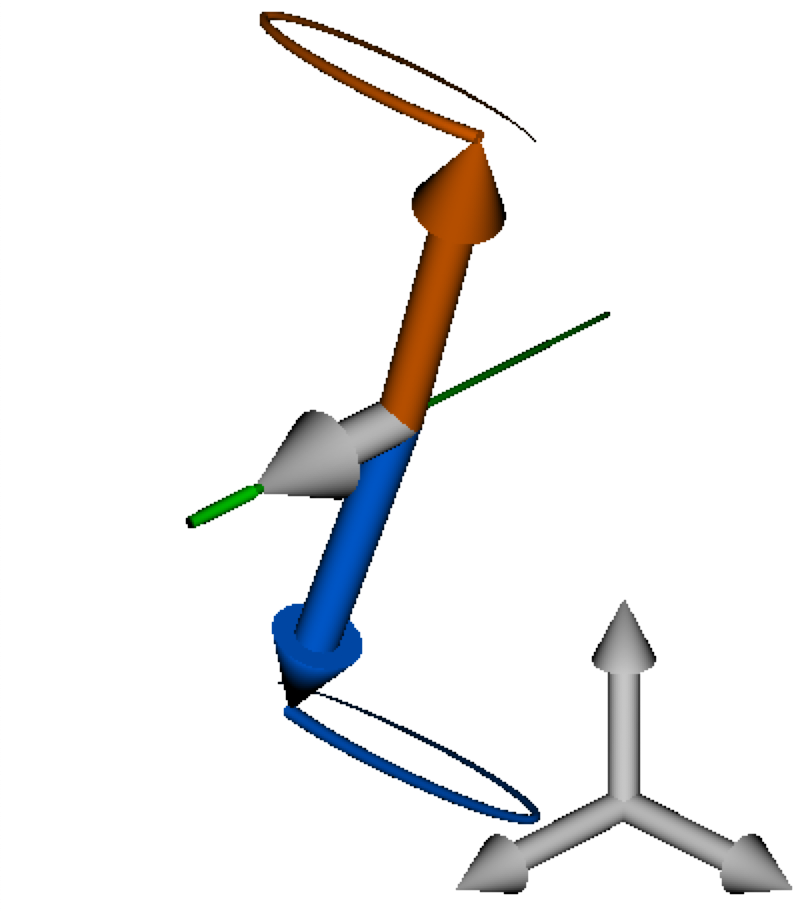}}
        \put( 1.9, 2.3){ \makebox(\boxsize,\boxsize){$\alpha$-mode} }
        \put(   2.2 , 0.2 ){\color{black}\scalebox{0.9}{\makebox{a}} }
        \put(   1.8 , 0.8 ){\color{black}\scalebox{0.9}{\makebox{b}} }
        \put(   0.95, 0.15){\color{black}\scalebox{0.9}{\makebox{c}} }
    }
    
    \put( 5.0, 0){
        \put( 0.0, 0){\includegraphics[width=2.3cm, trim={2.2cm 0cm -0.4cm 0cm},clip=true]{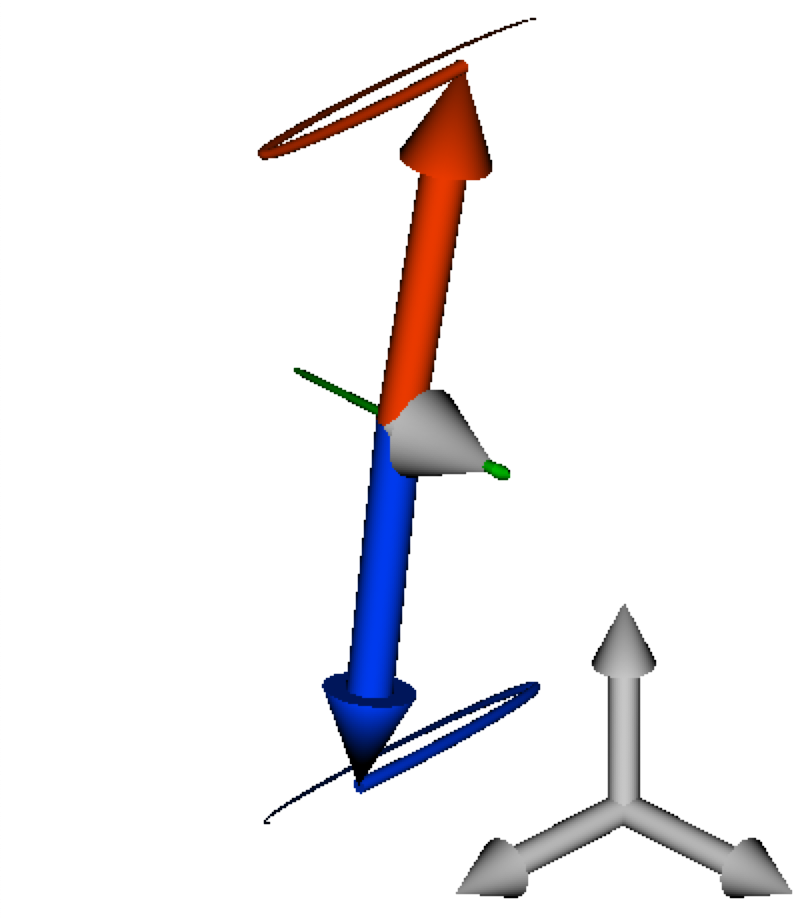}} 
        \put( 1.9, 2.3){ \makebox(\boxsize,\boxsize){$\beta$-mode} }
        \put(   2.2 , 0.2 ){\color{black}\scalebox{0.9}{\makebox{a}} }
        \put(   1.8 , 0.8 ){\color{black}\scalebox{0.9}{\makebox{b}} }
        \put(   0.95, 0.15){\color{black}\scalebox{0.9}{\makebox{c}} }
    }
  \end{picture}
  \caption{\label{fig:theory_double_gap}
  Adiabatic spin--wave energies of $\textnormal{Mn}_5\textnormal{Si}_3$ as a function of the external field applied along (a) the $c$--axis (perpendicular to the preferred axis but parallel to the second anisotropy easy--axis).
  The $\beta$--mode disperses with the field because it has linear polarization perpendicular to it.
  Meanwhile, the $\alpha$--mode has linear polarization parallel to the field (along the $c$--axis).
  (b) External field applied along the $a$--axis (parallel to the hard--axis).
  The field couples with the $\alpha$--mode which has linear polarization perpendicular to it.
  In this case, no energy crossing occurs. Lines in (a) and (b) are guides for the eyes.
  (c) and (d) precessional dynamics of the spins for biaxial anisotropy and without an external magnetic field for modes $\alpha$ and $\beta$, respectively.
  The red and blue arrows indicate antiparallel spins within a Mn2 triangle precessing around their equilibrium direction ($b$--axis).
  The central gray arrows represent the dynamics of the net magnetization of the system, which correspond to linear (longitudinal) precessions.
  }
\end{figure}
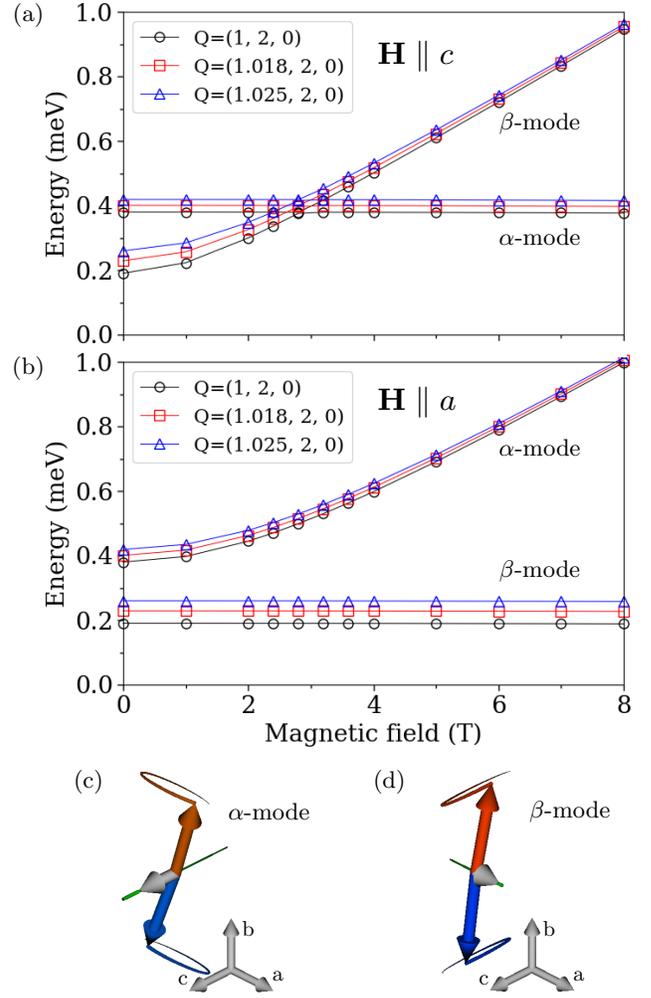

The results of our calculations for an external magnetic field applied along the $c$ crystal axis are shown in Fig.~\ref{fig:theory_double_gap}(a).
We observe that the energy of one mode is insensitive to the magnetic field ($\alpha$--mode), while the energy of the other mode ($\beta$--mode) increases monotonically for larger fields.
Moreover, the $\beta$--mode is lower in energy than the $\alpha$--mode for zero field, and for increasing field strength they eventually cross, which is in qualitative agreement with the INS data shown in Fig.~\ref{fig:exp_ine_spectra}(b).
No avoided crossing is observed with our model.
At zero field, the energy of the $\alpha$--mode is solely a function of $k^b$ and the magnetic exchange interactions.
The energy difference between the two modes is determined by $k^c$, which is zero for $k^c = 0$.
When $k^c = k^b$, the $\beta$--mode gap closes.
These results are made explicit by our analytical study of a corresponding one--dimensional antiferromagnetic system including nearest--neighbor--only exchange interaction $J$ and a biaxial anisotropy.
In the asymptotic limit of $k\ll J$~\cite{dos_santos_first-principles_2020}, the energies of two spin--wave modes are $E^\alpha\ \approx 2 S \sqrt{J k^b} $ and $E^\beta \approx 2 S \sqrt{J (k^b - k^c) }$.
We note that $k^c > k^b$ represents an unstable situation because the $c$--axis becomes the preferred one, but the magnetic moments are aligned along $b$.
If we apply the field along $a$, the field dependence of these modes inverts, as shown in Fig.~\ref{fig:theory_double_gap}(b).

\subsection{Precessional nature of the spin--wave modes}

To understand why the two spin--wave modes react so differently to the external magnetic field, we now discuss their precessional nature.
For zero field and $k^c=0$ (i.e. with uniaxial anisotropy), the modes are degenerate as pointed out before.
For these modes, the spins have an elliptical precession because in a certain instant of their revolution they are perfectly antiparallel but a quarter of revolution later, they are noncollinear~\cite{keffer_theory_1952}.
This noncollinear alignment is unfavored by the exchange interaction, making the precession elliptical.
The ellipses describing the precession of each mode have their major axes perpendicular to each other.
However, there is no preferred orientation for these major axes as long as the system remains isotropic in the $ac$--plane.
Besides, these modes can also be globally characterized by the dynamics of the net magnetization, which is defined by the sum of all spins at each instance of time.
For these modes, the net magnetization oscillates linearly along the minor ellipse axis of the corresponding mode, as illustrated by the central gray arrow in Fig.~\ref{fig:theory_double_gap}(c) and (d).
These modes are said to be linearly polarized along the oscillation axis of the net magnetization.
Furthermore, they have no net angular momentum and can be thought of as longitudinal modes~\cite{dos_santos_spin-resolved_2018}.

A preferred direction of the ellipse major axes can be achieved in two ways.
Firstly, we can apply a magnetic field perpendicularly to the preferred axis, i.e., $\perp b$--axis. 
Assuming that the field was applied along $c$, the $\alpha$--mode becomes linearly polarized along $c$, and consequently, the $\beta$--mode is polarized along $a$.
An oscillating net magnetization parallel to the field is not subjected to a torque due to that field.
Therefore, only the $\beta$--mode with polarization perpendicular to the magnetic field is affected such that its energy increases with the field.

Secondly, we can introduce a second anisotropy axis perpendicular to the first with weaker strength, let us say along $c$.
The energy of the mode with the major axis of the ellipse along $c$ ($\beta$--mode) is reduced because the system gains energy when the spins tilt in that direction.
The energy of the $\alpha$--mode, which tilts mostly perpendicularly to $c$, is not much affected.
Adding the magnetic field also along $c$ increases the energy of the $\beta$--mode as before, which eventually leads to the crossing of the energy of the two modes (see Figure~\ref{fig:theory_double_gap}(a)).
Figure~\ref{fig:theory_double_gap}(b) demonstrates that if we apply the magnetic field along $a$, therefore, perpendicular to the linear polarization of the $\alpha$--mode, then it would be the energy of this mode to increase with the field.

Polarized neutron scattering experiments around an AFM zone center capture the elliptic polarization of the magnon modes. Therefore, the proposed behavior regarding the polarization of the $\alpha$ and $\beta$--mode is also reflected in the polarized INS spectra shown in Fig.~\ref{fig:Capture5}. The ellipses major axis of the processional motion of the $\alpha$--mode is along the $a$--axis and consistently the intensity of this mode will appear in the subtracted spin fluctuations spectra obtained from LPA in $\langle{\delta}{M}_{a}\rangle$. Equivalently, the $\beta$--mode shows maximum intensity in $\langle{\delta}{M}_{c}\rangle$, since the major axis of the processional motion is along the $c$--axis.

\subsection{Canting of magnetic moments}

Polarized single--crystal neutron diffraction experiments~\cite{brown_antiferromagnetism_1995} in the AFM2 phase of $\textnormal{Mn}_5\textnormal{Si}_3$ showed that the magnetic moments are aligned along the $b$--axis and that there are no components of moments parallel to the $c$--axis. However, a more recent neutron diffraction study performed on polycrystalline samples~\cite{gottschilch_study_2012} suggested a deviation from the perfect collinearity, which is temperature--dependent increasing up to 8\,$^{\circ}$ near 70\,K with respect to the $b$--axis, but with the moments still confined in the $ab$--plane. Such a spin canting could originate from the Dzyaloshinskii--Moriya~\cite{Dzyaloshinsky_1958,Moriya_1960} exchange interaction (DMI). To investigate the field dependence of the two modes in the presence of DMI, we included the following term into our model Hamiltonian: 
\begin{equation}
\HH_\textnormal{DMI} = -\sum_{ij} \VEC D_{ij} \cdot (\VEC S_i \times \VEC S_j).
\end{equation}

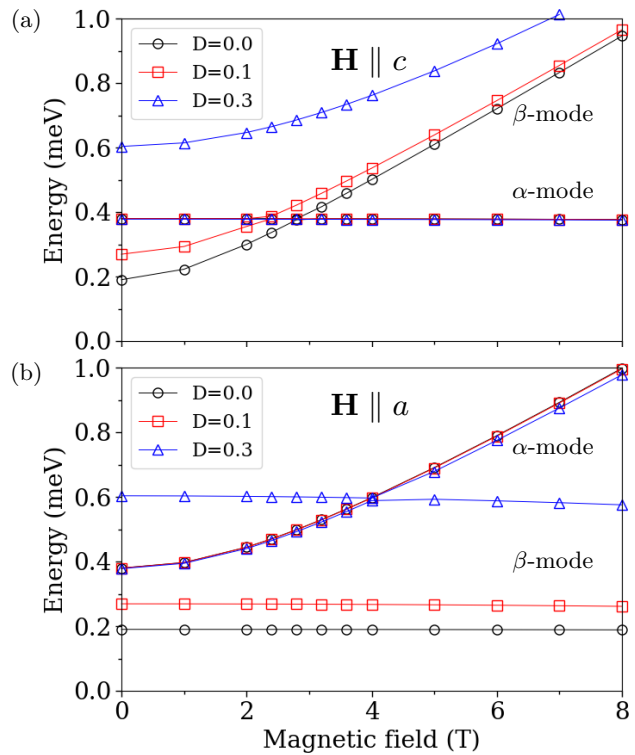
\begin{figure}[tb]
  \setlength{\unitlength}{1cm}
  \newcommand{\boxsize}{0.3}
  \begin{picture}(8,10.3)
    \put( 0.0, 4.65){ \includegraphics[width=8.0cm, trim={0cm 0cm 0cm 0cm},clip=true]{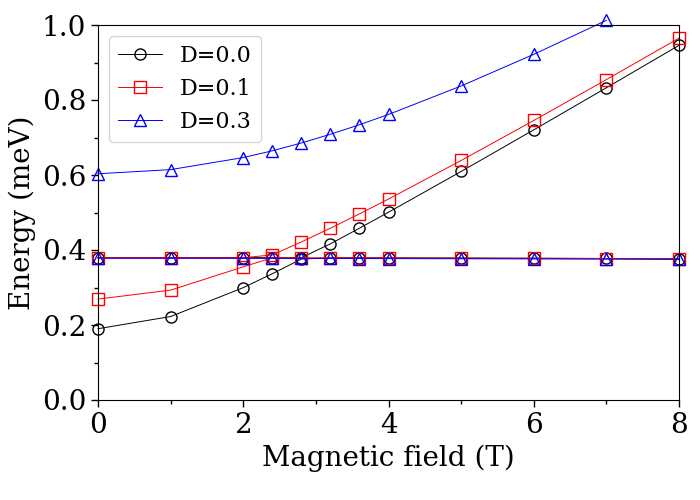} }
    \put( 0.0, 0.00){ \includegraphics[width=8.0cm, trim={0cm 0cm 0cm 0.1cm},clip=true]{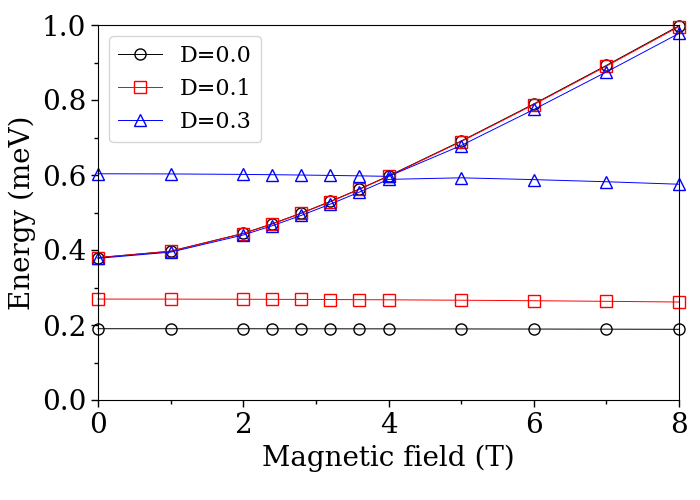} }
    
    \put(   4.0, 9.4 ){\color{black}\scalebox{1.3}{\makebox{$\VEC H \parallel c$}} }
    \put(   4.0, 4.8) {\color{black}\scalebox{1.3}{\makebox{$\VEC H \parallel a$}} }
    \put( 6.7, 7.65){ \makebox(\boxsize,\boxsize){$\alpha$-mode} }
    \put( 6.7, 8.65){ \makebox(\boxsize,\boxsize){$\beta$-mode} }
    \put(6.7, 4.25){ \makebox(\boxsize,\boxsize){$\alpha$-mode} }
    \put(6.7, 2.7){ \makebox(\boxsize,\boxsize){$\beta$-mode} }
    
    \linethickness{0.2cm}
    \put(1.1,5.6){ \color{white} \line(1,0){6.7}}
    
    \put(-0.3, 9.9){ \makebox(\boxsize,\boxsize){(a)} }
    \put(-0.3,  5.2){ \makebox(\boxsize,\boxsize){(b)} }

  \end{picture}
  \caption{\label{fig:theory_dispersion_with_DMI}
  Effects of the DMI on the spin--wave dispersion of $\textnormal{Mn}_5\textnormal{Si}_3$ at $\VEC Q = (1,2,0)$. 
  The external field was applied along (a) the $c$--axis (perpendicular to the preferred axis but parallel to the second anisotropy easy--axis), and along (b) the $a$--axis (parallel to the hard--axis).
  The DMI increases the energy of the $\beta$--mode and its effect on the $\alpha$--mode is almost negligible.
  $|\VEC D_{ij}| = D$ is in units of meV. Lines are guides for the eyes.
  }
\end{figure}
The results of the \textit{ab initio} calculation indicated that the DMI is indeed finite and smaller than 10\% of the exchange interaction for the Mn2 pair coupled by $J_1$ where $\VEC D_{ij} || \hat{\VEC c}$.
The impact of DMI on the $\alpha$ and $\beta$--mode for two field directions is shown in Fig.~\ref{fig:theory_dispersion_with_DMI}.
The effect of the DMI in the $\alpha$--mode is almost negligible, independently of the direction of the external magnetic field.
In contrast, the DMI causes the $\beta$--mode energy to increase.
For a field applied along the $a$--axis, the almost flat dispersion of the $\beta$--mode shifts to higher energies with increasing $D$, which eventually results in a crossing with the $\alpha$--mode (see the blue curves in  Fig.~\ref{fig:theory_dispersion_with_DMI}(b)).
When the field is applied along the $c$--axis, the increase in energy is accompanied by a change of the ``parabolic" dispersion of the $\beta$--mode, as shown in Fig.~\ref{fig:theory_dispersion_with_DMI}(a).
It is worth noticing that at zero fields, the energy of the $\alpha$--mode is solely given by the exchange interaction and the first anisotropy axis, while the $\beta$--mode energy is determined by the DMI in combination with the second anisotropy axis.
Based on our results from the INS measurements (see Fig.~\ref{fig:exp_ine_spectra}(b)), we can conclude since a crossing of the two modes is observed for $\VEC H\parallel \hat{\VEC c}$, that the DM exchange interaction should be very weak in the AFM2 phase of $\textnormal{Mn}_5\textnormal{Si}_3$.

\section{Conclusions}

We investigated the microscopic magnetic properties of the collinear antiferromagnetic phase of $\textnormal{Mn}_5\textnormal{Si}_3$ with inelastic neutron scattering measurements and density functional theory simulations.
The measurements have revealed two low--energy spin--wave modes that disperse with the wave--vector as expected for gapped antiferromagnetic magnons.
The high resolution of the measurements also allowed us to detect the small energy splitting between the two modes of about 0.2\,meV.
When a magnetic field was applied to the sample along the $c$--axis, the two modes showed a very different response to the field.
The energy of the lower--energy mode increased with increasing field, eventually crossing the energy of the other mode, which showed no observable response to the field.
These modes were further characterized by polarized neutron scattering measurements, which showed that their polarization is distinct and likely elliptical, with different major axes for each ellipse.
Our first--principles calculations capture the main features observed so far experimentally, namely, the collinear arrangement of the magnetic moments, the coexistence of magnetic and nonmagnetic Mn sites, and the biaxial magnetocrystalline anisotropy.
Also, we expanded the existing studies for collinear antiferromagnets with biaxial anisotropy regarding the polarization of the low energy spin--wave modes and their evolution under the influence of an external magnetic field.
Finally, we investigated theoretically the impact of the Dzyaloshinskii--Moriya interaction in the magnon spectrum and its implications depending on the strength of the anisotropy.
Our results could find echo in other similar systems where interesting magnonic phenomena occur in the field of antiferromagnetic spintronics.\vspace{20pt}

\section{Acknowledgments}

The neutron data collected at the Institut Laue Langevin for the present work are available at \onlinecite{data_IN12_NMF,data_IN12_MF,data_THALES_POL}.
N.B. acknowledges the support of JCNS through the Tasso Springer fellowship.
This work was also supported by the Brazilian funding agency CAPES under Project No. 13703/13-7 and the European Research Council (ERC) under the European Union's Horizon 2020 research and innovation program (ERC-consolidator Grant No. 681405-DYNASORE).
We gratefully acknowledge the computing time granted by JARA-HPC on the supercomputer JURECA at Forschungszentrum J\"ulich and by RWTH Aachen University.

\bibliography{paper_Mn5Si3}

\begin{thebibliography}{43}%
\makeatletter
\providecommand \@ifxundefined [1]{%
 \@ifx{#1\undefined}
}%
\providecommand \@ifnum [1]{%
 \ifnum #1\expandafter \@firstoftwo
 \else \expandafter \@secondoftwo
 \fi
}%
\providecommand \@ifx [1]{%
 \ifx #1\expandafter \@firstoftwo
 \else \expandafter \@secondoftwo
 \fi
}%
\providecommand \natexlab [1]{#1}%
\providecommand \enquote  [1]{``#1''}%
\providecommand \bibnamefont  [1]{#1}%
\providecommand \bibfnamefont [1]{#1}%
\providecommand \citenamefont [1]{#1}%
\providecommand \href@noop [0]{\@secondoftwo}%
\providecommand \href [0]{\begingroup \@sanitize@url \@href}%
\providecommand \@href[1]{\@@startlink{#1}\@@href}%
\providecommand \@@href[1]{\endgroup#1\@@endlink}%
\providecommand \@sanitize@url [0]{\catcode `\\12\catcode `\$12\catcode
  `\&12\catcode `\#12\catcode `\^12\catcode `\_12\catcode `\%12\relax}%
\providecommand \@@startlink[1]{}%
\providecommand \@@endlink[0]{}%
\providecommand \url  [0]{\begingroup\@sanitize@url \@url }%
\providecommand \@url [1]{\endgroup\@href {#1}{\urlprefix }}%
\providecommand \urlprefix  [0]{URL }%
\providecommand \Eprint [0]{\href }%
\providecommand \doibase [0]{http://dx.doi.org/}%
\providecommand \selectlanguage [0]{\@gobble}%
\providecommand \bibinfo  [0]{\@secondoftwo}%
\providecommand \bibfield  [0]{\@secondoftwo}%
\providecommand \translation [1]{[#1]}%
\providecommand \BibitemOpen [0]{}%
\providecommand \bibitemStop [0]{}%
\providecommand \bibitemNoStop [0]{.\EOS\space}%
\providecommand \EOS [0]{\spacefactor3000\relax}%
\providecommand \BibitemShut  [1]{\csname bibitem#1\endcsname}%
\let\auto@bib@innerbib\@empty
\bibitem [{\citenamefont {Baltz}\ \emph {et~al.}(2018)\citenamefont {Baltz},
  \citenamefont {Manchon}, \citenamefont {Tsoi}, \citenamefont {Moriyama},
  \citenamefont {Ono},\ and\ \citenamefont {Tserkovnyak}}]{Baltz_2018}%
  \BibitemOpen
  \bibfield  {author} {\bibinfo {author} {\bibfnamefont {V.}~\bibnamefont
  {Baltz}}, \bibinfo {author} {\bibfnamefont {A.}~\bibnamefont {Manchon}},
  \bibinfo {author} {\bibfnamefont {M.}~\bibnamefont {Tsoi}}, \bibinfo {author}
  {\bibfnamefont {T.}~\bibnamefont {Moriyama}}, \bibinfo {author}
  {\bibfnamefont {T.}~\bibnamefont {Ono}}, \ and\ \bibinfo {author}
  {\bibfnamefont {Y.}~\bibnamefont {Tserkovnyak}},\ }\href {\doibase
  10.1103/RevModPhys.90.015005} {\bibfield  {journal} {\bibinfo  {journal}
  {Rev. Mod. Phys.}\ }\textbf {\bibinfo {volume} {90}},\ \bibinfo {pages}
  {015005} (\bibinfo {year} {2018})}\BibitemShut {NoStop}%
\bibitem [{\citenamefont {Brown}\ \emph {et~al.}(1992)\citenamefont {Brown},
  \citenamefont {Forsyth}, \citenamefont {Nunez},\ and\ \citenamefont
  {Tasset}}]{brown_low-temperature_1992}%
  \BibitemOpen
  \bibfield  {author} {\bibinfo {author} {\bibfnamefont {P.~J.}\ \bibnamefont
  {Brown}}, \bibinfo {author} {\bibfnamefont {J.~B.}\ \bibnamefont {Forsyth}},
  \bibinfo {author} {\bibfnamefont {V.}~\bibnamefont {Nunez}}, \ and\ \bibinfo
  {author} {\bibfnamefont {F.}~\bibnamefont {Tasset}},\ }\href {\doibase
  10.1088/0953-8984/4/49/029} {\bibfield  {journal} {\bibinfo  {journal}
  {Journal of Physics: Condensed Matter}\ }\textbf {\bibinfo {volume} {4}},\
  \bibinfo {pages} {10025} (\bibinfo {year} {1992})}\BibitemShut {NoStop}%
\bibitem [{\citenamefont {Sürgers}\ \emph {et~al.}(2014)\citenamefont
  {Sürgers}, \citenamefont {Fischer}, \citenamefont {Winkel},\ and\
  \citenamefont {Löhneysen}}]{surgers_large_2014}%
  \BibitemOpen
  \bibfield  {author} {\bibinfo {author} {\bibfnamefont {C.}~\bibnamefont
  {Sürgers}}, \bibinfo {author} {\bibfnamefont {G.}~\bibnamefont {Fischer}},
  \bibinfo {author} {\bibfnamefont {P.}~\bibnamefont {Winkel}}, \ and\ \bibinfo
  {author} {\bibfnamefont {H.~v.}\ \bibnamefont {Löhneysen}},\ }\href
  {\doibase 10.1038/ncomms4400} {\bibfield  {journal} {\bibinfo  {journal}
  {Nature Communications}\ }\textbf {\bibinfo {volume} {5}},\ \bibinfo {pages}
  {3400} (\bibinfo {year} {2014})}\BibitemShut {NoStop}%
\bibitem [{\citenamefont {Biniskos}\ \emph {et~al.}(2018)\citenamefont
  {Biniskos}, \citenamefont {Schmalzl}, \citenamefont {Raymond}, \citenamefont
  {Petit}, \citenamefont {Steffens}, \citenamefont {Persson},\ and\
  \citenamefont {Brückel}}]{biniskos_spin_2018}%
  \BibitemOpen
  \bibfield  {author} {\bibinfo {author} {\bibfnamefont {N.}~\bibnamefont
  {Biniskos}}, \bibinfo {author} {\bibfnamefont {K.}~\bibnamefont {Schmalzl}},
  \bibinfo {author} {\bibfnamefont {S.}~\bibnamefont {Raymond}}, \bibinfo
  {author} {\bibfnamefont {S.}~\bibnamefont {Petit}}, \bibinfo {author}
  {\bibfnamefont {P.}~\bibnamefont {Steffens}}, \bibinfo {author}
  {\bibfnamefont {J.}~\bibnamefont {Persson}}, \ and\ \bibinfo {author}
  {\bibfnamefont {T.}~\bibnamefont {Brückel}},\ }\href {\doibase
  10.1103/PhysRevLett.120.257205} {\bibfield  {journal} {\bibinfo  {journal}
  {Physical Review Letters}\ }\textbf {\bibinfo {volume} {120}},\ \bibinfo
  {pages} {257205} (\bibinfo {year} {2018})}\BibitemShut {NoStop}%
\bibitem [{\citenamefont {Das}\ \emph {et~al.}(2016)\citenamefont {Das},
  \citenamefont {Balasubramanian}, \citenamefont {Manchanda}, \citenamefont
  {Mukherjee}, \citenamefont {Skomski}, \citenamefont {Hadjipanayis},\ and\
  \citenamefont {Sellmyer}}]{das_mn5si3_2016}%
  \BibitemOpen
  \bibfield  {author} {\bibinfo {author} {\bibfnamefont {B.}~\bibnamefont
  {Das}}, \bibinfo {author} {\bibfnamefont {B.}~\bibnamefont
  {Balasubramanian}}, \bibinfo {author} {\bibfnamefont {P.}~\bibnamefont
  {Manchanda}}, \bibinfo {author} {\bibfnamefont {P.}~\bibnamefont
  {Mukherjee}}, \bibinfo {author} {\bibfnamefont {R.}~\bibnamefont {Skomski}},
  \bibinfo {author} {\bibfnamefont {G.~C.}\ \bibnamefont {Hadjipanayis}}, \
  and\ \bibinfo {author} {\bibfnamefont {D.~J.}\ \bibnamefont {Sellmyer}},\
  }\href {\doibase 10.1021/acs.nanolett.5b04360} {\bibfield  {journal}
  {\bibinfo  {journal} {Nano Letters}\ }\textbf {\bibinfo {volume} {16}},\
  \bibinfo {pages} {1132} (\bibinfo {year} {2016})},\ \bibinfo {note}
  {publisher: American Chemical Society}\BibitemShut {NoStop}%
\bibitem [{\citenamefont {Sun}\ \emph {et~al.}(2020)\citenamefont {Sun},
  \citenamefont {Sun}, \citenamefont {He}, \citenamefont {Yang},\ and\
  \citenamefont {Wang}}]{sun_millimeters_2020}%
  \BibitemOpen
  \bibfield  {author} {\bibinfo {author} {\bibfnamefont {Y.}~\bibnamefont
  {Sun}}, \bibinfo {author} {\bibfnamefont {B.}~\bibnamefont {Sun}}, \bibinfo
  {author} {\bibfnamefont {J.}~\bibnamefont {He}}, \bibinfo {author}
  {\bibfnamefont {G.}~\bibnamefont {Yang}}, \ and\ \bibinfo {author}
  {\bibfnamefont {C.}~\bibnamefont {Wang}},\ }\href {\doibase
  10.1038/s41467-019-14244-5} {\bibfield  {journal} {\bibinfo  {journal}
  {Nature Communications}\ }\textbf {\bibinfo {volume} {11}},\ \bibinfo {pages}
  {647} (\bibinfo {year} {2020})},\ \bibinfo {note} {number: 1 Publisher:
  Nature Publishing Group}\BibitemShut {NoStop}%
\bibitem [{\citenamefont {Gottschilch}\ \emph {et~al.}(2012)\citenamefont
  {Gottschilch}, \citenamefont {Gourdon}, \citenamefont {Persson},
  \citenamefont {Cruz}, \citenamefont {Petricek},\ and\ \citenamefont
  {Brueckel}}]{gottschilch_study_2012}%
  \BibitemOpen
  \bibfield  {author} {\bibinfo {author} {\bibfnamefont {M.}~\bibnamefont
  {Gottschilch}}, \bibinfo {author} {\bibfnamefont {O.}~\bibnamefont
  {Gourdon}}, \bibinfo {author} {\bibfnamefont {J.}~\bibnamefont {Persson}},
  \bibinfo {author} {\bibfnamefont {C.~d.~l.}\ \bibnamefont {Cruz}}, \bibinfo
  {author} {\bibfnamefont {V.}~\bibnamefont {Petricek}}, \ and\ \bibinfo
  {author} {\bibfnamefont {T.}~\bibnamefont {Brueckel}},\ }\href {\doibase
  10.1039/C2JM00154C} {\bibfield  {journal} {\bibinfo  {journal} {Journal of
  Materials Chemistry}\ }\textbf {\bibinfo {volume} {22}},\ \bibinfo {pages}
  {15275} (\bibinfo {year} {2012})},\ \bibinfo {note} {publisher: The Royal
  Society of Chemistry}\BibitemShut {NoStop}%
\bibitem [{\citenamefont {Brown}\ and\ \citenamefont
  {Forsyth}(1995)}]{brown_antiferromagnetism_1995}%
  \BibitemOpen
  \bibfield  {author} {\bibinfo {author} {\bibfnamefont {P.~J.}\ \bibnamefont
  {Brown}}\ and\ \bibinfo {author} {\bibfnamefont {J.~B.}\ \bibnamefont
  {Forsyth}},\ }\href {\doibase 10.1088/0953-8984/7/39/004} {\bibfield
  {journal} {\bibinfo  {journal} {Journal of Physics: Condensed Matter}\
  }\textbf {\bibinfo {volume} {7}},\ \bibinfo {pages} {7619} (\bibinfo {year}
  {1995})}\BibitemShut {NoStop}%
\bibitem [{\citenamefont {Sürgers}\ \emph {et~al.}(2017)\citenamefont
  {Sürgers}, \citenamefont {Wolf}, \citenamefont {Adelmann}, \citenamefont
  {Kittler}, \citenamefont {Fischer},\ and\ \citenamefont
  {Löhneysen}}]{surgers_switching_2017}%
  \BibitemOpen
  \bibfield  {author} {\bibinfo {author} {\bibfnamefont {C.}~\bibnamefont
  {Sürgers}}, \bibinfo {author} {\bibfnamefont {T.}~\bibnamefont {Wolf}},
  \bibinfo {author} {\bibfnamefont {P.}~\bibnamefont {Adelmann}}, \bibinfo
  {author} {\bibfnamefont {W.}~\bibnamefont {Kittler}}, \bibinfo {author}
  {\bibfnamefont {G.}~\bibnamefont {Fischer}}, \ and\ \bibinfo {author}
  {\bibfnamefont {H.~v.}\ \bibnamefont {Löhneysen}},\ }\href {\doibase
  10.1038/srep42982} {\bibfield  {journal} {\bibinfo  {journal} {Scientific
  Reports}\ }\textbf {\bibinfo {volume} {7}},\ \bibinfo {pages} {42982}
  (\bibinfo {year} {2017})}\BibitemShut {NoStop}%
\bibitem [{\citenamefont {Vinokurova}\ \emph {et~al.}(1990)\citenamefont
  {Vinokurova}, \citenamefont {Ivanov}, \citenamefont {Kulatov},\ and\
  \citenamefont {Vlasov}}]{vinokurova_magnetic_1990}%
  \BibitemOpen
  \bibfield  {author} {\bibinfo {author} {\bibfnamefont {L.}~\bibnamefont
  {Vinokurova}}, \bibinfo {author} {\bibfnamefont {V.}~\bibnamefont {Ivanov}},
  \bibinfo {author} {\bibfnamefont {E.}~\bibnamefont {Kulatov}}, \ and\
  \bibinfo {author} {\bibfnamefont {A.}~\bibnamefont {Vlasov}},\ }\href
  {\doibase 10.1016/S0304-8853(10)80040-X} {\bibfield  {journal} {\bibinfo
  {journal} {Journal of Magnetism and Magnetic Materials}\ }\textbf {\bibinfo
  {volume} {90-91}},\ \bibinfo {pages} {121} (\bibinfo {year}
  {1990})}\BibitemShut {NoStop}%
\bibitem [{\citenamefont {Luccas}\ \emph {et~al.}(2019)\citenamefont {Luccas},
  \citenamefont {Sánchez-Santolino}, \citenamefont {Correa-Orellana},
  \citenamefont {Mompean}, \citenamefont {García-Hernández},\ and\
  \citenamefont {Suderow}}]{luccas_magnetic_2019}%
  \BibitemOpen
  \bibfield  {author} {\bibinfo {author} {\bibfnamefont {R.~F.}\ \bibnamefont
  {Luccas}}, \bibinfo {author} {\bibfnamefont {G.}~\bibnamefont
  {Sánchez-Santolino}}, \bibinfo {author} {\bibfnamefont {A.}~\bibnamefont
  {Correa-Orellana}}, \bibinfo {author} {\bibfnamefont {F.~J.}\ \bibnamefont
  {Mompean}}, \bibinfo {author} {\bibfnamefont {M.}~\bibnamefont
  {García-Hernández}}, \ and\ \bibinfo {author} {\bibfnamefont
  {H.}~\bibnamefont {Suderow}},\ }\href {\doibase 10.1016/j.jmmm.2019.165451}
  {\bibfield  {journal} {\bibinfo  {journal} {Journal of Magnetism and Magnetic
  Materials}\ }\textbf {\bibinfo {volume} {489}},\ \bibinfo {pages} {165451}
  (\bibinfo {year} {2019})}\BibitemShut {NoStop}%
\bibitem [{\citenamefont {Das}\ \emph {et~al.}(2019)\citenamefont {Das},
  \citenamefont {Mandal}, \citenamefont {Dutta}, \citenamefont {Pramanick},\
  and\ \citenamefont {Chatterjee}}]{das_observation_2019}%
  \BibitemOpen
  \bibfield  {author} {\bibinfo {author} {\bibfnamefont {S.~C.}\ \bibnamefont
  {Das}}, \bibinfo {author} {\bibfnamefont {K.}~\bibnamefont {Mandal}},
  \bibinfo {author} {\bibfnamefont {P.}~\bibnamefont {Dutta}}, \bibinfo
  {author} {\bibfnamefont {S.}~\bibnamefont {Pramanick}}, \ and\ \bibinfo
  {author} {\bibfnamefont {S.}~\bibnamefont {Chatterjee}},\ }\href {\doibase
  10.1103/PhysRevB.100.024409} {\bibfield  {journal} {\bibinfo  {journal}
  {Physical Review B}\ }\textbf {\bibinfo {volume} {100}},\ \bibinfo {pages}
  {024409} (\bibinfo {year} {2019})},\ \bibinfo {note} {publisher: American
  Physical Society}\BibitemShut {NoStop}%
\bibitem [{\citenamefont {Silva}\ \emph {et~al.}(2002)\citenamefont {Silva},
  \citenamefont {Brown},\ and\ \citenamefont {Forsyth}}]{silva_magnetic_2002}%
  \BibitemOpen
  \bibfield  {author} {\bibinfo {author} {\bibfnamefont {M.~R.}\ \bibnamefont
  {Silva}}, \bibinfo {author} {\bibfnamefont {P.~J.}\ \bibnamefont {Brown}}, \
  and\ \bibinfo {author} {\bibfnamefont {J.~B.}\ \bibnamefont {Forsyth}},\
  }\href {\doibase 10.1088/0953-8984/14/37/307} {\bibfield  {journal} {\bibinfo
   {journal} {Journal of Physics: Condensed Matter}\ }\textbf {\bibinfo
  {volume} {14}},\ \bibinfo {pages} {8707} (\bibinfo {year} {2002})},\ \bibinfo
  {note} {publisher: IOP Publishing}\BibitemShut {NoStop}%
\bibitem [{\citenamefont {Al-Kanani}\ and\ \citenamefont
  {Booth}(1995)}]{Alkanani_1995}%
  \BibitemOpen
  \bibfield  {author} {\bibinfo {author} {\bibfnamefont {H.}~\bibnamefont
  {Al-Kanani}}\ and\ \bibinfo {author} {\bibfnamefont {J.}~\bibnamefont
  {Booth}},\ }\href {\doibase https://doi.org/10.1016/0304-8853(94)01157-5}
  {\bibfield  {journal} {\bibinfo  {journal} {Journal of Magnetism and Magnetic
  Materials}\ }\textbf {\bibinfo {volume} {140-144}},\ \bibinfo {pages} {1539 }
  (\bibinfo {year} {1995})},\ \bibinfo {note} {international Conference on
  Magnetism}\BibitemShut {NoStop}%
\bibitem [{\citenamefont {Irizawa}\ \emph {et~al.}(2002)\citenamefont
  {Irizawa}, \citenamefont {Yamasaki}, \citenamefont {Okazaki}, \citenamefont
  {Kasai}, \citenamefont {Sekiyama}, \citenamefont {Imada}, \citenamefont
  {Suga}, \citenamefont {Kulatov}, \citenamefont {Ohta},\ and\ \citenamefont
  {Nanba}}]{Irizawa_2002}%
  \BibitemOpen
  \bibfield  {author} {\bibinfo {author} {\bibfnamefont {A.}~\bibnamefont
  {Irizawa}}, \bibinfo {author} {\bibfnamefont {A.}~\bibnamefont {Yamasaki}},
  \bibinfo {author} {\bibfnamefont {M.}~\bibnamefont {Okazaki}}, \bibinfo
  {author} {\bibfnamefont {S.}~\bibnamefont {Kasai}}, \bibinfo {author}
  {\bibfnamefont {A.}~\bibnamefont {Sekiyama}}, \bibinfo {author}
  {\bibfnamefont {S.}~\bibnamefont {Imada}}, \bibinfo {author} {\bibfnamefont
  {S.}~\bibnamefont {Suga}}, \bibinfo {author} {\bibfnamefont {E.}~\bibnamefont
  {Kulatov}}, \bibinfo {author} {\bibfnamefont {H.}~\bibnamefont {Ohta}}, \
  and\ \bibinfo {author} {\bibfnamefont {T.}~\bibnamefont {Nanba}},\ }\href
  {\doibase https://doi.org/10.1016/S0038-1098(02)00475-1} {\bibfield
  {journal} {\bibinfo  {journal} {Solid State Communications}\ }\textbf
  {\bibinfo {volume} {124}},\ \bibinfo {pages} {1 } (\bibinfo {year}
  {2002})}\BibitemShut {NoStop}%
\bibitem [{\citenamefont {Sürgers}\ \emph {et~al.}(2016)\citenamefont
  {Sürgers}, \citenamefont {Kittler}, \citenamefont {Wolf},\ and\
  \citenamefont {Löhneysen}}]{Suergers_2016}%
  \BibitemOpen
  \bibfield  {author} {\bibinfo {author} {\bibfnamefont {C.}~\bibnamefont
  {Sürgers}}, \bibinfo {author} {\bibfnamefont {W.}~\bibnamefont {Kittler}},
  \bibinfo {author} {\bibfnamefont {T.}~\bibnamefont {Wolf}}, \ and\ \bibinfo
  {author} {\bibfnamefont {H.~v.}\ \bibnamefont {Löhneysen}},\ }\href
  {\doibase 10.1063/1.4943759} {\bibfield  {journal} {\bibinfo  {journal} {AIP
  Advances}\ }\textbf {\bibinfo {volume} {6}},\ \bibinfo {pages} {055604}
  (\bibinfo {year} {2016})},\ \Eprint
  {http://arxiv.org/abs/https://doi.org/10.1063/1.4943759}
  {https://doi.org/10.1063/1.4943759} \BibitemShut {NoStop}%
\bibitem [{\citenamefont {Songlin}\ \emph {et~al.}(2002)\citenamefont
  {Songlin}, \citenamefont {Tegus}, \citenamefont {Brück}, \citenamefont
  {Klaasse}, \citenamefont {de~Boer},\ and\ \citenamefont
  {Buschow}}]{songlin_magnetic_2002}%
  \BibitemOpen
  \bibfield  {author} {\bibinfo {author} {\bibfnamefont {D.}~\bibnamefont
  {Songlin}}, \bibinfo {author} {\bibfnamefont {O.}~\bibnamefont {Tegus}},
  \bibinfo {author} {\bibfnamefont {E.}~\bibnamefont {Brück}}, \bibinfo
  {author} {\bibfnamefont {J.~C.~P.}\ \bibnamefont {Klaasse}}, \bibinfo
  {author} {\bibfnamefont {F.~R.}\ \bibnamefont {de~Boer}}, \ and\ \bibinfo
  {author} {\bibfnamefont {K.~H.~J.}\ \bibnamefont {Buschow}},\ }\href
  {\doibase 10.1016/S0925-8388(01)01776-5} {\bibfield  {journal} {\bibinfo
  {journal} {Journal of Alloys and Compounds}\ }\textbf {\bibinfo {volume}
  {334}},\ \bibinfo {pages} {249} (\bibinfo {year} {2002})}\BibitemShut
  {NoStop}%
\bibitem [{\citenamefont {Schmalzl}\ \emph {et~al.}(2016)\citenamefont
  {Schmalzl}, \citenamefont {Schmidt}, \citenamefont {Raymond}, \citenamefont
  {Feilbach}, \citenamefont {Mounier}, \citenamefont {Vettard},\ and\
  \citenamefont {Brückel}}]{schmalzl_upgrade_2016}%
  \BibitemOpen
  \bibfield  {author} {\bibinfo {author} {\bibfnamefont {K.}~\bibnamefont
  {Schmalzl}}, \bibinfo {author} {\bibfnamefont {W.}~\bibnamefont {Schmidt}},
  \bibinfo {author} {\bibfnamefont {S.}~\bibnamefont {Raymond}}, \bibinfo
  {author} {\bibfnamefont {H.}~\bibnamefont {Feilbach}}, \bibinfo {author}
  {\bibfnamefont {C.}~\bibnamefont {Mounier}}, \bibinfo {author} {\bibfnamefont
  {B.}~\bibnamefont {Vettard}}, \ and\ \bibinfo {author} {\bibfnamefont
  {T.}~\bibnamefont {Brückel}},\ }\href {\doibase 10.1016/j.nima.2016.02.067}
  {\bibfield  {journal} {\bibinfo  {journal} {Nuclear Instruments and Methods
  in Physics Research Section A: Accelerators, Spectrometers, Detectors and
  Associated Equipment}\ }\textbf {\bibinfo {volume} {819}},\ \bibinfo {pages}
  {89} (\bibinfo {year} {2016})}\BibitemShut {NoStop}%
\bibitem [{\citenamefont {Regnault}\ \emph {et~al.}(2004)\citenamefont
  {Regnault}, \citenamefont {Geffray}, \citenamefont {Fouilloux}, \citenamefont
  {Longuet}, \citenamefont {Mantegazza}, \citenamefont {Tasset}, \citenamefont
  {Lelièvre-Berna}, \citenamefont {Pujol}, \citenamefont {Bourgeat-Lami},
  \citenamefont {Kernavanois}, \citenamefont {Thomas},\ and\ \citenamefont
  {Gibert}}]{regnault_spherical_2004}%
  \BibitemOpen
  \bibfield  {author} {\bibinfo {author} {\bibfnamefont {L.~P.}\ \bibnamefont
  {Regnault}}, \bibinfo {author} {\bibfnamefont {B.}~\bibnamefont {Geffray}},
  \bibinfo {author} {\bibfnamefont {P.}~\bibnamefont {Fouilloux}}, \bibinfo
  {author} {\bibfnamefont {B.}~\bibnamefont {Longuet}}, \bibinfo {author}
  {\bibfnamefont {F.}~\bibnamefont {Mantegazza}}, \bibinfo {author}
  {\bibfnamefont {F.}~\bibnamefont {Tasset}}, \bibinfo {author} {\bibfnamefont
  {E.}~\bibnamefont {Lelièvre-Berna}}, \bibinfo {author} {\bibfnamefont
  {S.}~\bibnamefont {Pujol}}, \bibinfo {author} {\bibfnamefont
  {E.}~\bibnamefont {Bourgeat-Lami}}, \bibinfo {author} {\bibfnamefont
  {N.}~\bibnamefont {Kernavanois}}, \bibinfo {author} {\bibfnamefont
  {M.}~\bibnamefont {Thomas}}, \ and\ \bibinfo {author} {\bibfnamefont
  {Y.}~\bibnamefont {Gibert}},\ }\href {\doibase 10.1016/j.physb.2004.03.211}
  {\bibfield  {journal} {\bibinfo  {journal} {Physica B: Condensed Matter}\
  }\bibinfo {series} {Proceedings of the {Third} {European} {Conference} on
  {Neutron} {Scattering}},\ \textbf {\bibinfo {volume} {350}},\ \bibinfo
  {pages} {E811} (\bibinfo {year} {2004})}\BibitemShut {NoStop}%
\bibitem [{\citenamefont {Chatterji}(2006)}]{chatterji_neutron_2006}%
  \BibitemOpen
  \bibinfo {editor} {\bibfnamefont {T.}~\bibnamefont {Chatterji}},\ ed.,\
  \href@noop {} {\emph {\bibinfo {title} {Neutron scattering from magnetic
  materials}}},\ \bibinfo {edition} {1st}\ ed.\ (\bibinfo  {publisher}
  {Elsevier},\ \bibinfo {address} {Amsterdam ; Boston},\ \bibinfo {year}
  {2006})\ \bibinfo {note} {oCLC: ocm61748467}\BibitemShut {NoStop}%
\bibitem [{com()}]{comment}%
  \BibitemOpen
  \href@noop {} {}\bibinfo {note} {In this article we use the orthorhombic
  coordinate system and the scattering vector $\VEC Q$ is expressed in
  Cartesian coordinates $\VEC Q = (Q_{h}, Q_{k}, Q_{l})$ given in reciprocal
  lattice units (r.l.u.). The wave--vector $\VEC q$ is related to the momentum
  transfer through $\hbar\VEC Q = \hbar\VEC G + \hbar\VEC q$, where $\VEC G$ is
  a Brillouin zone center and $\VEC G = (h, k, l)$.}\BibitemShut {Stop}%
\bibitem [{\citenamefont {Ibuka}\ \emph {et~al.}(2017)\citenamefont {Ibuka},
  \citenamefont {Itoh}, \citenamefont {Yokoo},\ and\ \citenamefont
  {Endoh}}]{Ibuka}%
  \BibitemOpen
  \bibfield  {author} {\bibinfo {author} {\bibfnamefont {S.}~\bibnamefont
  {Ibuka}}, \bibinfo {author} {\bibfnamefont {S.}~\bibnamefont {Itoh}},
  \bibinfo {author} {\bibfnamefont {T.}~\bibnamefont {Yokoo}}, \ and\ \bibinfo
  {author} {\bibfnamefont {Y.}~\bibnamefont {Endoh}},\ }\href {\doibase
  10.1103/PhysRevB.95.224406} {\bibfield  {journal} {\bibinfo  {journal} {Phys.
  Rev. B}\ }\textbf {\bibinfo {volume} {95}},\ \bibinfo {pages} {224406}
  (\bibinfo {year} {2017})}\BibitemShut {NoStop}%
\bibitem [{\citenamefont {Papanikolaou}\ \emph {et~al.}(2002)\citenamefont
  {Papanikolaou}, \citenamefont {Zeller},\ and\ \citenamefont
  {Dederichs}}]{papanikolaou_conceptual_2002}%
  \BibitemOpen
  \bibfield  {author} {\bibinfo {author} {\bibfnamefont {N.}~\bibnamefont
  {Papanikolaou}}, \bibinfo {author} {\bibfnamefont {R.}~\bibnamefont
  {Zeller}}, \ and\ \bibinfo {author} {\bibfnamefont {P.~H.}\ \bibnamefont
  {Dederichs}},\ }\href {\doibase 10.1088/0953-8984/14/11/304} {\bibfield
  {journal} {\bibinfo  {journal} {Journal of Physics: Condensed Matter}\
  }\textbf {\bibinfo {volume} {14}},\ \bibinfo {pages} {2799} (\bibinfo {year}
  {2002})},\ \bibinfo {note} {publisher: IOP Publishing}\BibitemShut {NoStop}%
\bibitem [{\citenamefont {Vosko}\ \emph {et~al.}(1980)\citenamefont {Vosko},
  \citenamefont {Wilk},\ and\ \citenamefont {Nusair}}]{Vosko1980}%
  \BibitemOpen
  \bibfield  {author} {\bibinfo {author} {\bibfnamefont {S.}~\bibnamefont
  {Vosko}}, \bibinfo {author} {\bibfnamefont {L.}~\bibnamefont {Wilk}}, \ and\
  \bibinfo {author} {\bibfnamefont {M.}~\bibnamefont {Nusair}},\ }\href
  {\doibase 10.1139/p80-159} {\bibfield  {journal} {\bibinfo  {journal} {Can.
  J. Phys.}\ }\textbf {\bibinfo {volume} {58}},\ \bibinfo {pages} {1200}
  (\bibinfo {year} {1980})}\BibitemShut {NoStop}%
\bibitem [{\citenamefont {Wildberger}\ \emph {et~al.}(1995)\citenamefont
  {Wildberger}, \citenamefont {Lang}, \citenamefont {Zeller},\ and\
  \citenamefont {Dederichs}}]{wildberger_fermi-dirac_1995}%
  \BibitemOpen
  \bibfield  {author} {\bibinfo {author} {\bibfnamefont {K.}~\bibnamefont
  {Wildberger}}, \bibinfo {author} {\bibfnamefont {P.}~\bibnamefont {Lang}},
  \bibinfo {author} {\bibfnamefont {R.}~\bibnamefont {Zeller}}, \ and\ \bibinfo
  {author} {\bibfnamefont {P.~H.}\ \bibnamefont {Dederichs}},\ }\href {\doibase
  10.1103/PhysRevB.52.11502} {\bibfield  {journal} {\bibinfo  {journal}
  {Physical Review B}\ }\textbf {\bibinfo {volume} {52}},\ \bibinfo {pages}
  {11502} (\bibinfo {year} {1995})},\ \bibinfo {note} {publisher: American
  Physical Society}\BibitemShut {NoStop}%
\bibitem [{\citenamefont {Liechtenstein}\ \emph {et~al.}(1987)\citenamefont
  {Liechtenstein}, \citenamefont {Katsnelson}, \citenamefont {Antropov},\ and\
  \citenamefont {Gubanov}}]{liechtenstein_local_1987}%
  \BibitemOpen
  \bibfield  {author} {\bibinfo {author} {\bibfnamefont {A.~I.}\ \bibnamefont
  {Liechtenstein}}, \bibinfo {author} {\bibfnamefont {M.~I.}\ \bibnamefont
  {Katsnelson}}, \bibinfo {author} {\bibfnamefont {V.~P.}\ \bibnamefont
  {Antropov}}, \ and\ \bibinfo {author} {\bibfnamefont {V.~A.}\ \bibnamefont
  {Gubanov}},\ }\href {\doibase 10.1016/0304-8853(87)90721-9} {\bibfield
  {journal} {\bibinfo  {journal} {Journal of Magnetism and Magnetic Materials}\
  }\textbf {\bibinfo {volume} {67}},\ \bibinfo {pages} {65} (\bibinfo {year}
  {1987})}\BibitemShut {NoStop}%
\bibitem [{\citenamefont {Ebert}\ and\ \citenamefont
  {Mankovsky}(2009)}]{Ebert2009}%
  \BibitemOpen
  \bibfield  {author} {\bibinfo {author} {\bibfnamefont {H.}~\bibnamefont
  {Ebert}}\ and\ \bibinfo {author} {\bibfnamefont {S.}~\bibnamefont
  {Mankovsky}},\ }\href {http://prb.aps.org/abstract/PRB/v79/i4/e045209}
  {\bibfield  {journal} {\bibinfo  {journal} {Phys. Rev. B}\ }\textbf {\bibinfo
  {volume} {79}},\ \bibinfo {pages} {045209} (\bibinfo {year}
  {2009})}\BibitemShut {NoStop}%
\bibitem [{\citenamefont {dos Santos}\ \emph {et~al.}(2018)\citenamefont {dos
  Santos}, \citenamefont {dos Santos~Dias}, \citenamefont {Guimarães},
  \citenamefont {Bouaziz},\ and\ \citenamefont
  {Lounis}}]{dos_santos_spin-resolved_2018}%
  \BibitemOpen
  \bibfield  {author} {\bibinfo {author} {\bibfnamefont {F.~J.}\ \bibnamefont
  {dos Santos}}, \bibinfo {author} {\bibfnamefont {M.}~\bibnamefont {dos
  Santos~Dias}}, \bibinfo {author} {\bibfnamefont {F.~S.~M.}\ \bibnamefont
  {Guimarães}}, \bibinfo {author} {\bibfnamefont {J.}~\bibnamefont {Bouaziz}},
  \ and\ \bibinfo {author} {\bibfnamefont {S.}~\bibnamefont {Lounis}},\ }\href
  {\doibase 10.1103/PhysRevB.97.024431} {\bibfield  {journal} {\bibinfo
  {journal} {Physical Review B}\ }\textbf {\bibinfo {volume} {97}},\ \bibinfo
  {pages} {024431} (\bibinfo {year} {2018})}\BibitemShut {NoStop}%
\bibitem [{\citenamefont {Holstein}\ and\ \citenamefont
  {Primakoff}(1940)}]{holstein_field_1940}%
  \BibitemOpen
  \bibfield  {author} {\bibinfo {author} {\bibfnamefont {T.}~\bibnamefont
  {Holstein}}\ and\ \bibinfo {author} {\bibfnamefont {H.}~\bibnamefont
  {Primakoff}},\ }\href {\doibase 10.1103/PhysRev.58.1098} {\bibfield
  {journal} {\bibinfo  {journal} {Physical Review}\ }\textbf {\bibinfo {volume}
  {58}},\ \bibinfo {pages} {1098} (\bibinfo {year} {1940})}\BibitemShut
  {NoStop}%
\bibitem [{\citenamefont {Bogoljubov}(1958)}]{bogoljubov_new_1958}%
  \BibitemOpen
  \bibfield  {author} {\bibinfo {author} {\bibfnamefont {N.~N.}\ \bibnamefont
  {Bogoljubov}},\ }\href {\doibase 10.1007/BF02745585} {\bibfield  {journal}
  {\bibinfo  {journal} {Il Nuovo Cimento (1955-1965)}\ }\textbf {\bibinfo
  {volume} {7}},\ \bibinfo {pages} {794} (\bibinfo {year} {1958})}\BibitemShut
  {NoStop}%
\bibitem [{\citenamefont {dos Santos}\ \emph
  {et~al.}(2020{\natexlab{a}})\citenamefont {dos Santos}, \citenamefont {dos
  Santos~Dias},\ and\ \citenamefont {Lounis}}]{dos_santos_nonreciprocity_2020}%
  \BibitemOpen
  \bibfield  {author} {\bibinfo {author} {\bibfnamefont {F.~J.}\ \bibnamefont
  {dos Santos}}, \bibinfo {author} {\bibfnamefont {M.}~\bibnamefont {dos
  Santos~Dias}}, \ and\ \bibinfo {author} {\bibfnamefont {S.}~\bibnamefont
  {Lounis}},\ }\href {\doibase 10.1103/PhysRevB.102.104401} {\bibfield
  {journal} {\bibinfo  {journal} {Physical Review B}\ }\textbf {\bibinfo
  {volume} {102}},\ \bibinfo {pages} {104401} (\bibinfo {year}
  {2020}{\natexlab{a}})},\ \bibinfo {note} {publisher: American Physical
  Society}\BibitemShut {NoStop}%
\bibitem [{\citenamefont {dos Santos}\ \emph
  {et~al.}(2020{\natexlab{b}})\citenamefont {dos Santos}, \citenamefont {dos
  Santos~Dias},\ and\ \citenamefont {Lounis}}]{dos_santos_modeling_2020}%
  \BibitemOpen
  \bibfield  {author} {\bibinfo {author} {\bibfnamefont {F.~J.}\ \bibnamefont
  {dos Santos}}, \bibinfo {author} {\bibfnamefont {M.}~\bibnamefont {dos
  Santos~Dias}}, \ and\ \bibinfo {author} {\bibfnamefont {S.}~\bibnamefont
  {Lounis}},\ }\href {\doibase 10.1103/PhysRevB.102.104436} {\bibfield
  {journal} {\bibinfo  {journal} {Physical Review B}\ }\textbf {\bibinfo
  {volume} {102}},\ \bibinfo {pages} {104436} (\bibinfo {year}
  {2020}{\natexlab{b}})},\ \bibinfo {note} {publisher: American Physical
  Society}\BibitemShut {NoStop}%
\bibitem [{\citenamefont {dos Santos}\ \emph {et~al.}(2017)\citenamefont {dos
  Santos}, \citenamefont {dos Santos~Dias},\ and\ \citenamefont
  {Lounis}}]{dos_santos_first-principles_2017}%
  \BibitemOpen
  \bibfield  {author} {\bibinfo {author} {\bibfnamefont {F.~J.}\ \bibnamefont
  {dos Santos}}, \bibinfo {author} {\bibfnamefont {M.}~\bibnamefont {dos
  Santos~Dias}}, \ and\ \bibinfo {author} {\bibfnamefont {S.}~\bibnamefont
  {Lounis}},\ }\href {\doibase 10.1103/PhysRevB.95.134408} {\bibfield
  {journal} {\bibinfo  {journal} {Physical Review B}\ }\textbf {\bibinfo
  {volume} {95}},\ \bibinfo {pages} {134408} (\bibinfo {year} {2017})},\
  \bibinfo {note} {publisher: American Physical Society}\BibitemShut {NoStop}%
\bibitem [{\citenamefont {Mendive-Tapia}\ and\ \citenamefont
  {Staunton}(2019)}]{MendiveTapia2019}%
  \BibitemOpen
  \bibfield  {author} {\bibinfo {author} {\bibfnamefont {E.}~\bibnamefont
  {Mendive-Tapia}}\ and\ \bibinfo {author} {\bibfnamefont {J.~B.}\ \bibnamefont
  {Staunton}},\ }\href {\doibase 10.1103/PhysRevB.99.144424} {\bibfield
  {journal} {\bibinfo  {journal} {Phys. Rev. B}\ }\textbf {\bibinfo {volume}
  {99}},\ \bibinfo {pages} {144424} (\bibinfo {year} {2019})}\BibitemShut
  {NoStop}%
\bibitem [{\citenamefont {Maraytta}\ \emph {et~al.}(2020)\citenamefont
  {Maraytta}, \citenamefont {Voigt}, \citenamefont {Salazar~Mejía},
  \citenamefont {Friese}, \citenamefont {Skourski}, \citenamefont {Perßon},
  \citenamefont {Salman},\ and\ \citenamefont {Brückel}}]{Maraytta2020}%
  \BibitemOpen
  \bibfield  {author} {\bibinfo {author} {\bibfnamefont {N.}~\bibnamefont
  {Maraytta}}, \bibinfo {author} {\bibfnamefont {J.}~\bibnamefont {Voigt}},
  \bibinfo {author} {\bibfnamefont {C.}~\bibnamefont {Salazar~Mejía}},
  \bibinfo {author} {\bibfnamefont {K.}~\bibnamefont {Friese}}, \bibinfo
  {author} {\bibfnamefont {Y.}~\bibnamefont {Skourski}}, \bibinfo {author}
  {\bibfnamefont {J.}~\bibnamefont {Perßon}}, \bibinfo {author} {\bibfnamefont
  {S.~M.}\ \bibnamefont {Salman}}, \ and\ \bibinfo {author} {\bibfnamefont
  {T.}~\bibnamefont {Brückel}},\ }\href {\doibase 10.1063/5.0020780}
  {\bibfield  {journal} {\bibinfo  {journal} {Journal of Applied Physics}\
  }\textbf {\bibinfo {volume} {128}},\ \bibinfo {pages} {103903} (\bibinfo
  {year} {2020})},\ \Eprint
  {http://arxiv.org/abs/https://doi.org/10.1063/5.0020780}
  {https://doi.org/10.1063/5.0020780} \BibitemShut {NoStop}%
\bibitem [{\citenamefont {Keffer}\ and\ \citenamefont
  {Kittel}(1952)}]{keffer_theory_1952}%
  \BibitemOpen
  \bibfield  {author} {\bibinfo {author} {\bibfnamefont {F.}~\bibnamefont
  {Keffer}}\ and\ \bibinfo {author} {\bibfnamefont {C.}~\bibnamefont
  {Kittel}},\ }\href {\doibase 10.1103/PhysRev.85.329} {\bibfield  {journal}
  {\bibinfo  {journal} {Physical Review}\ }\textbf {\bibinfo {volume} {85}},\
  \bibinfo {pages} {329} (\bibinfo {year} {1952})},\ \bibinfo {note}
  {publisher: American Physical Society}\BibitemShut {NoStop}%
\bibitem [{\citenamefont {Rezende}\ \emph {et~al.}(2019)\citenamefont
  {Rezende}, \citenamefont {Azevedo},\ and\ \citenamefont
  {Rodríguez-Suárez}}]{rezende_introduction_2019}%
  \BibitemOpen
  \bibfield  {author} {\bibinfo {author} {\bibfnamefont {S.~M.}\ \bibnamefont
  {Rezende}}, \bibinfo {author} {\bibfnamefont {A.}~\bibnamefont {Azevedo}}, \
  and\ \bibinfo {author} {\bibfnamefont {R.~L.}\ \bibnamefont
  {Rodríguez-Suárez}},\ }\href {\doibase 10.1063/1.5109132} {\bibfield
  {journal} {\bibinfo  {journal} {Journal of Applied Physics}\ }\textbf
  {\bibinfo {volume} {126}},\ \bibinfo {pages} {151101} (\bibinfo {year}
  {2019})},\ \bibinfo {note} {publisher: American Institute of
  Physics}\BibitemShut {NoStop}%
\bibitem [{\citenamefont {dos
  Santos}(2020)}]{dos_santos_first-principles_2020}%
  \BibitemOpen
  \bibfield  {author} {\bibinfo {author} {\bibfnamefont {F.~J.}\ \bibnamefont
  {dos Santos}},\ }\emph {\bibinfo {title} {First-principles study of
  collective spin excitations in noncollinear magnets}},\ \href {\doibase
  10.18154/RWTH-2020-01879} {Ph.D. thesis},\ \bibinfo  {school}
  {Forschungszentrum Jülich GmbH, Zentralbibliothek, Verlag} (\bibinfo {year}
  {2020}),\ \bibinfo {note} {iSBN: 9783958064591 Number:
  RWTH-2020-01879}\BibitemShut {NoStop}%
\bibitem [{\citenamefont {Dzyaloshinsky}(1958)}]{Dzyaloshinsky_1958}%
  \BibitemOpen
  \bibfield  {author} {\bibinfo {author} {\bibfnamefont {I.}~\bibnamefont
  {Dzyaloshinsky}},\ }\href {\doibase
  https://doi.org/10.1016/0022-3697(58)90076-3} {\bibfield  {journal} {\bibinfo
   {journal} {Journal of Physics and Chemistry of Solids}\ }\textbf {\bibinfo
  {volume} {4}},\ \bibinfo {pages} {241 } (\bibinfo {year} {1958})}\BibitemShut
  {NoStop}%
\bibitem [{\citenamefont {Moriya}(1960)}]{Moriya_1960}%
  \BibitemOpen
  \bibfield  {author} {\bibinfo {author} {\bibfnamefont {T.}~\bibnamefont
  {Moriya}},\ }\href {\doibase 10.1103/PhysRev.120.91} {\bibfield  {journal}
  {\bibinfo  {journal} {Phys. Rev.}\ }\textbf {\bibinfo {volume} {120}},\
  \bibinfo {pages} {91} (\bibinfo {year} {1960})}\BibitemShut {NoStop}%
\bibitem [{dat(2016)}]{data_IN12_NMF}%
  \BibitemOpen
  \href {https://doi.ill.fr/10.5291/ILL-DATA.CRG-2331} {\bibfield  {journal}
  {\bibinfo  {journal} {https://doi.ill.fr/10.5291/ILL-DATA.CRG-2331}\ }
  (\bibinfo {year} {2016})}\BibitemShut {NoStop}%
\bibitem [{dat(2019{\natexlab{a}})}]{data_IN12_MF}%
  \BibitemOpen
  \href {https://doi.ill.fr/10.5291/ILL-DATA.CRG-2620} {\bibfield  {journal}
  {\bibinfo  {journal} {https://doi.ill.fr/10.5291/ILL-DATA.CRG-2620}\ }
  (\bibinfo {year} {2019}{\natexlab{a}})}\BibitemShut {NoStop}%
\bibitem [{dat(2019{\natexlab{b}})}]{data_THALES_POL}%
  \BibitemOpen
  \href {https://doi.ill.fr/10.5291/ILL-DATA.4-01-1618} {\bibfield  {journal}
  {\bibinfo  {journal} {https://doi.ill.fr/10.5291/ILL-DATA.4-01-1618}\ }
  (\bibinfo {year} {2019}{\natexlab{b}})}\BibitemShut {NoStop}%
\end{thebibliography}%

\end{document}